\documentclass[lettersize,journal]{IEEEtran}

\usepackage{amsmath,amsfonts,graphics,amssymb,epsfig,subfigure,color,cite}
\usepackage{amsthm,textcomp}
\theoremstyle{plain}
\newtheoremstyle{mystyle}%                % Name
  {0mm}%                                  % Space above
  {0mm}%                                  % Space below
  {}%                                     % Body font
  {4mm}%                                  % Indent amount
  {\bfseries}%                            % Theorem head font
  {:}%                                    % Punctuation after theorem head
  { }%                                    % Space after theorem head, ' ', or \newline
  {\thmname{#1}\thmnumber{ #2}\thmnote{ (#3)}}%                                     % Theorem head spec (can be left empty, meaning `normal')
\theoremstyle{mystyle}
\usepackage{bm}

\usepackage{array}
\usepackage{multirow}
\usepackage{enumerate}
\usepackage{algorithm}
\usepackage{algpseudocode}
\usepackage{algcompatible}
\algblockdefx{WHILE}{ENDWHILE}[1]%
  {\textbf{while }#1 \textbf{do}}%
  {\textbf{end}}
\algblockdefx{FORP}{ENDFORP}[1]%
  {\textbf{for }#1 \textbf{do in parallel}}%
  {\textbf{end}}
\algblockdefx{FOR}{ENDFOR}[1]%
  {\textbf{for }#1 \textbf{do}}%
  {\textbf{end}}
\algblockdefx{noEnd}{noEndEnd}[1]
    {#1}
    {{\Statex}}
\algblockdefx{BULLET}{EndBULLET}{$\bullet$ }{}
\algnotext{EndBULLET}
\algblockdefx{PROC}{ENDPROC}[1]
    {\textbf{procedure }#1 {}}
    {\textbf{end procedure}}
\algblockdefx{IF}{ENDIF}[1]%
  {\textbf{if }#1 \textbf{then}}%
  {\textbf{end}}
\algnewcommand\algorithmicprocedure{\textbf{procedure}}
\algnewcommand\FUNC{\item[\algorithmicprocedure]}%
\algnewcommand\algorithmicendprocedure{\textbf{end procedure}}
\algnewcommand\ENDFUNC{\item[\algorithmicendprocedure]}%
 \usepackage{setspace}
\let\Algorithm\algorithm
\renewcommand\algorithm[1][]{\Algorithm[#1]\setstretch{1.4}}
\usepackage{float}
\usepackage{makecell}
\usepackage{graphicx}
\usepackage{xcolor}
\usepackage{soul}

\newtheorem{thm}{Theorem}
\newtheorem{lem}{Lemma}

\newtheorem{prop}{Proposition}

\usepackage{bbm}

\newcommand{\argmin}{\operatornamewithlimits{argmin}}

\makeatletter
\newcommand{\vast}{\bBigg@{4.5}}
\newcommand{\Vast}{\bBigg@{7.5}}
\makeatother

\allowdisplaybreaks

\begin{document}
    \title{SplitMAC: Wireless Split Learning over Multiple Access Channels}
    % \title{Communication Efficient Split Learning over Wireless Multiple Access Channels}
	%\title{FedQCS: Federated Learning via \\  Quantized Compressed Sensing} <- For TSP,
	%\title{Quantized Compressed Sensing for Communication-Efficient Federated Learning}
	\author{Seonjung Kim, Yongjeong Oh, \IEEEmembership{Student Member,~IEEE}, and Yo-Seb Jeon, \IEEEmembership{Member,~IEEE}
	    %\thanks{This work was supported in part by the National Research Foundation of Korea (NRF) grant funded by the Korea government (MSIT) (No. 2022R1C1C1010074), and in part by the U.S. National Science Foundation under Grant CNS-2114267.} 
	    \thanks{Seonjung Kim, Yongjeong Oh, and Yo-Seb Jeon are with the Department of Electrical Engineering, POSTECH, Pohang, Gyeongbuk 37673, Republic of Korea (e-mails: \{seonjung.kim, yongjeongoh, yoseb.jeon\}@postech.ac.kr).}
	    %Yongjeong Oh, Namyoon Lee, Yo-Seb Jeon, and H. Vincent Poor
	    %\thanks{This paper will be presented in part at the 2021 IEEE Global Communications Conference Workshops \cite{Conference}.}
		%\thanks{Y. Oh and Y.-S. Jeon are with the Department of Electrical Engineering, POSTECH, Pohang, Gyeongbuk 37673, South Korea (e-mails: \{yongjeongoh,nylee,yoseb.jeon\}@postech.ac.kr).}
		%\thanks{Mingzhe Chen is with the Department of Electrical and Computer Engineering and Institute for Data Science and Computing, University of Miami, Coral Gables, FL 33146, USA (e-mail: mingzhe.chen@miami.edu).}
		%\thanks{Walid Saad is with the Wireless@VT, Bradley Department of Electrical and Computer Engineering, Virginia Tech, Arlington, VA 22203, USA (e-mail: walids@vt.edu).}
		%\thanks{H. V. Poor is with the Department of Electrical Engineering, Princeton University, Princeton, NJ 08544 (e-mail: poor@princeton.edu).}
		%This work was supported in part by Samsung Research Funding $\&$ Incubation Center of Samsung Electronics under Project Number SRFC-IT1702-00, and in part by the National Science Foundation under Grant No. NSF-CCF-1527079. This work was presented in part at the 2018 IEEE 87th Vehicular Techonology Conference (VTC2018-Spring).}
	}
	\vspace{-2mm}	
	
	\maketitle
	\vspace{-12mm}

	\begin{abstract} % up to 200 words
         This paper presents a novel split learning (SL) framework, referred to as SplitMAC, which reduces the latency of SL by leveraging simultaneous uplink transmission over multiple access channels. The key strategy is to divide devices into multiple groups and allow the devices within the same group to simultaneously transmit their smashed data and device-side models over the multiple access channels. The optimization problem of device grouping to minimize SL latency is formulated, and the benefit of device grouping in reducing the uplink latency of SL is theoretically derived. By examining a two-device grouping case, two asymptotically-optimal algorithms are devised for device grouping in low and high signal-to-noise ratio (SNR) scenarios, respectively, while providing proofs of their optimality. By merging these algorithms, a near-optimal device grouping algorithm is proposed to cover a wide range of SNR. Our SL framework is also extended to consider practical fading channels and to support a general group size. Simulation results demonstrate that our SL framework with the proposed device grouping algorithm is superior to existing SL frameworks in reducing SL latency.  
        % This paper presents a novel split learning (SL) framework, referred to as SplitMAC, which reduces the latency of SL by leveraging simultaneous uplink transmission over multiple access channels. The key strategy is to divide devices into multiple groups and allow the devices within the same group to simultaneously transmit their smashed data and device-side models over the multiple access channels. The benefit of device grouping in reducing the uplink latency of SL is theoretically derived based on the capacity region of the multiple access channel. The optimization problem of device grouping to minimize SL latency is formulated. By focusing on a two-user grouping case, optimal device grouping algorithms for low signal-to-noise ratio (SNR) and high SNR scenarios are devised, respectively. By merging these algorithms, a near-optimal device grouping algorithm is proposed to cover a wide range of SNR values. Simulation results demonstrate that our SL framework with the proposed device grouping algorithm is superior to existing SL frameworks in reducing SL latency.  
	% 
        \end{abstract}

	\begin{IEEEkeywords}
		Split learning, multiple access channel, device grouping, device clustering, non-orthogonal multiple access 
	\end{IEEEkeywords}

	\section{Introduction}\label{Sec:Intro}

    %The deployment of Internet of Things (IoT) devices generates a huge amount of distributed data increasingly. The main technology that actively utilizes these data is machine learning (ML), especially deep neural networks (DNN). The major challenge incurred by traditional centralized learning (CL) is the privacy issue since CL requires the collection of the raw data of each IoT device. To overcome the issue, distributed deep learning (DDL) that trains the global model by aggregating locally trained models is investigated thanks to its privacy-enhancing property \cite{dean2012large}. Federated learning (FL) is considered a promising solution that requires only the gradient of each local model \cite{Konecny:15,mcmahan17a,niknam2020federated,jeon2020compressive}. Despite its advantage, the training procedure of FL incurs significant computational costs and storage requirements on devices, especially when training large-scale models.
    
    The proliferation of Internet of Things (IoT) devices has led to an exponential surge in distributed data generation. One pivotal technology that effectively harnesses this data is machine learning (ML), notably deep neural networks (DNNs). However, the conventional approach of centralized learning faces practical challenges due to data privacy and communication overhead, as centralized learning requires the collection of raw data from each IoT device. 
    Federated learning (FL) has emerged as a prospective solution to address these challenges, requiring solely gradients or model updates from individual local models \cite{Konecny:15,mcmahan17a,niknam2020federated}. 
    FL also offers reduced data transmission by processing data locally and exchanging only model parameters or gradients, conserving bandwidth, especially beneficial for devices with large data volumes \cite{Konecny:15}. Additionally, FL utilizes edge computing resources, distributing computational load and potentially reducing dependence on centralized data centers \cite{mcmahan17a}.
        % distributed learning has been explored, which is a methodology that refines a global model by aggregating locally trained models. This approach holds promise for enhancing privacy \cite{dean2012large}. Among these methods, federated learning (FL) has emerged as a prospective solution, requiring solely the gradient data from individual local models \cite{Konecny:15,mcmahan17a,niknam2020federated,jeon2020compressive}. 
%\textcolor{red}{Additionally, the reduced data transmission and utilization of edge computing motivates the usage of FL.} 
    Despite these advantages, the training process of FL imposes notable computational and storage burdens on devices, particularly when dealing with large-scale models \cite{jeon2020compressive,FedSQCS}.

    Recently, split learning (SL) has attracted increasing attention as a promising remedy to address the limitations of FL, while providing the advantages of FL such as enhancing data privacy and utilizing edge computing \cite{SL,vepakomma2018split}.
    Within a typical SL framework, the global model is partitioned into two distinct sub-models: the device-side model and the server-side model. Each device hosts the device-side model, encompassing the initial layers of the global model, while the parameter server maintains the server-side model, comprising the remaining layers. The SL training process unfolds in a round-robin manner across the devices. In each training iteration, individual devices execute forward propagation (FP) on their device-side models, utilizing a mini-batch that exclusively resides on the device. This yields intermediate activation values known as {\em smashed data}. Subsequently, this smashed data is transmitted to the server, which then continues the forward propagation process on the server-side model, utilizing the received smashed data. Following the completion of the forward propagation process for both models, the server initiates the backward propagation (BP) on the server-side model, leading to the derivation of an intermediate gradient. This gradient is conveyed back to the device, where the backward propagation process continues on the device-side model. With the conclusion of the backward propagation process for both-side models, gradients are computed, consequently facilitating the adjustment of the global model. This iterative sequence is then replicated for the subsequent device in line. This framework can also be adapted for scenarios involving vertically distributed data, known as vertical FL \cite{VFL}.

    SL offers a promising alternative to FL, particularly for large-scale model training, as it allows devices to share the burden of the global model with a powerful central server, thereby reducing computational costs and storage needs on devices \cite{letaief2021edge, tran2022privacy}. More specifically, the investigation in \cite{SL_FL_comparison} demonstrated that SL is more communication-efficient than FL, particularly as the model's complexity increases. This efficiency is achieved by offloading computational workloads to the resource-rich server. Additionally, the study in \cite{SL_convergence} showed that sequential SL is more robust to data heterogeneity than a typical FL framework \cite{mcmahan17a}, resulting in lower error bounds and faster convergence. Despite its advantages, implementing SL in practical applications is challenging due to significant communication delays \cite{letaief2021edge,tran2022privacy}. These delays stem from the frequent transmission of smashed data from devices and intermediate gradients from the server, increasing not only with the number of devices but also with the size of the smashed data or the intermediate gradients.

    %However, communication remains a significant challenge in distributed learning \cite{li2014communication}, even within the SL framework. This is because the SL training process requires frequent transmission of smashed data among devices, causing significant communication delays that increase with the number of the devices as well as the size of the smashed data.
    
    %\subsection{Prior Works}
    %To address the challenge of communication latency in SL, various communication-efficient SL frameworks have been studied. One promising solution is to reduce the communication overhead. In \cite{xu2022adaptive}, authors proposed adaptive model compression for reduction of the workload of FL. In \cite{SL_SQ_1}, the authors leveraged the top-$S$ sparsification technique for smashed data and intermediate gradient. In \cite{SL_SQ_2}, the authors proposed a quantization technique in which the smashed data is quantized based on $K$-means clustering. However, these approaches fundamentally lower the precision and accompany calculation errors during learning.

    To alleviate the communication latency of SL, various communication-efficient SL frameworks were developed in the literature \cite{shiranthika2023decentralized}. 
    %To address the challenge caused by the communication latency in SL, researchers have developed various communication-efficient SL frameworks.
    One of the representative approaches toward this direction is to compress the smashed data and/or intermediate gradients, in order to directly minimize the SL's communication overhead. 
    %One particularly promising avenue involves minimizing communication overhead. For instance, in \cite{xu2022adaptive}, the authors introduced an adaptive model compression method to alleviate the workload associated with FL. 
    For example, in \cite{SL_SQ_1}, a communication-efficient SL framework was developed by leveraging the top-$S$ sparsification technique for both smashed data and intermediate gradients. Similarly, in \cite{SL_SQ_2}, a quantization technique was proposed, wherein the smashed data undergoes quantization using $K$-means clustering. In \cite{oh2023communication}, adaptive feature-wise dropout was applied to the intermediate features. Subsequently, the communication overhead was further reduced by the adaptive feature-wise quantization. Despite their potential benefits, these approaches inherently compromise precision, introducing calculation errors during the training process. 

    %In fact, the substantial communication latency of SL stems from its sequential learning manner. Split federated learning (SFL) \cite{thapa2022splitfed} is a breakthrough that parallelizes the device-side models' learning process. Motivated by this parallel framework, parallel split learning (PSL) \cite{CPSL} parallelizes the communication between devices and PS using frequency division multiple access (FDMA).  Also, the authors in \cite{EPSL} proposed a parallel SL framework using FDMA and reduced the communication overhead via last-layer gradient aggregation. In \cite{CPSL} and \cite{EPSL}, authors optimized the subchannel allocation using greedy-based algorithms. The high computational complexity of greedy-based algorithms may harm the learning latency. 

    Another promising approach for realizing communication-efficient SL is to parallelize the SL training process across different devices. 
    %The considerable communication latency inherent in split learning (SL) is primarily attributed to its sequential learning approach. A significant breakthrough in this regard is represented by 
    A significant breakthrough in this regard is a {\em parallel} training approach considered in \cite{thapa2022splitfed}. In this approach, the devices update their device-side models in parallel, while the AP aggregates these models to construct an updated device-side model. This approach, however, may slow down the convergence speed of SL due to independently training and aggregating different device-side models, especially when the datasets of different devices are not identically distributed.
    To overcome the limitation of the parallel training, a {\em cluster-wise} training approach was adopted in \cite{CPSL,EPSL}, in which devices are divided into multiple clusters, and the model update is executed sequentially across the clusters. In particular, the authors in  \cite{CPSL,EPSL} employ frequency division multiple access (FDMA) to realize the parallel training process among the devices within the cluster.
    %A similar SL framework was also introduced in \cite{EPSL} \textcolor{red}{by utilizing FDMA and reduced the communication overhead via last-layer gradient aggregation.} 
    %Additionally, the authors in \cite{EPSL} proposed a parallel SL framework using FDMA and reduced the communication overhead via last-layer gradient aggregation. 
    A common limitation of the cluster-wise training based on FDMA is the requirement for non-overlapping frequency bands to avoid interference among different devices. This leads to an inefficient utilization of the frequency resources, which in turn increases the communication latency. 
    %\textcolor{blue}{Morevoer, both \cite{CPSL} and \cite{EPSL} undertook optimization of subchannel allocation using greedy-based algorithms. The high computational complexity of greedy-based algorithms may harm the total learning latency.}
    It is well-known that non-orthogonal multiple access (NOMA) provides better utilization of the available frequency resources than orthogonal multiple access such as FDMA, by supporting simultaneous transmission of multiple devices using the same frequency resources \cite{chen2017exploiting,vaezi2019interplay,wei2019performance}. Additionally, in \cite{chen2017optimization}, the superiority of NOMA over orthogonal multiple access (OMA) was proved in the sense of sum rate under optimal resource allocation. Despite these advantages, none of the existing studies have explored an SL framework that can leverage the advantage of NOMA for reducing the communication latency of SL.    

    To fill this research gap, this paper presents a novel SL framework that alleviates the latency of SL by harnessing the advantage of NOMA for the first time. In our framework, devices are divided into multiple groups, and the devices within the same group simultaneously transmit their data over wireless multiple access channels (MACs). An optimization problem for device grouping is formulated, and its solution is analyzed to prove the advantage of the device grouping for reducing the SL latency. Device grouping algorithms are also developed for a two-device grouping case, whose optimality can be guaranteed under certain signal-to-noise ratio (SNR) conditions. %hrough these contributions, this paper not only pioneers the exploration of NOMA in the context of SL, but also offers a practical way of maximizing the benefit of NOMA in latency reduction. 
    %The presented framework not only explores the potential of NOMA in the context of SL for the first time, but also maximizes the benefit of NOMA based on a latent analysis. 
    The major contributions of this paper are summarized below.
    \begin{itemize}
        \item We propose a novel SL framework, referred to as SL over multiple access channels (SplitMAC), designed to reduce the latency of SL. The core strategy of SplitMAC is to divide devices into multiple groups and allow the devices within the same group to simultaneously transmit their smashed data and device-side models over the multiple access channels. In addition to this strategy, SpitMAC further reduces the latency of SL by employing additional latency-reduction strategies: (i) local updates of server-side model and (ii) simultaneous uplink-downlink transmission, which have not been considered in the literature. %Specifically, we focus on scenarios where only two devices are grouped together.
        
        \item We formulate an optimization problem to determine the best device grouping strategy that minimizes the uplink latency of SL. 
        %\textcolor{blue}{Unlike the previous studies on latency minimization in NOMA system represented in terms of predetermined transmission rates \cite{task_offloading, noh2021delay}, we formulate it as the minimum latency by delving into which specific rates within the feasible region minimize latency.}
        We then derive the necessary and sufficient conditions on transmission rates to minimize the latency by analyzing the solution of the optimization problem. Based on these results, we theoretically prove the advantage of the device grouping for reducing the uplink latency in SplitMAC, compared to the case without device grouping. 
        %Its objective function forms a sub-problem which minimizes the uplink latency for a given group. 
        % We find a closed-form solution to this sub-problem. Using this solution, 
        Unlike the previous studies which represent the latency in terms of predetermined transmission rates \cite{task_offloading, noh2021delay}, we formulate it as the minimum latency by investigating the characteristics of the optimal rates in the feasible region that can minimize the latency.
    
        \item We develop device grouping algorithms for reducing the uplink latency of SplitMAC by focusing on a two-device grouping case. To be more specific, we devise two asymptotically-optimal device grouping algorithms and prove their optimality under high-SNR and low-SNR conditions, respectively. 
        We then develop a near-optimal device grouping algorithm to cover a wide range of SNR, while leveraging the advantages of these two algorithms.
        %SNR-specific grouping algorithms which are asymptotically optimal in specific scenarios and prove their optimalities, respectively. 

        \item Through simulations, we demonstrate the superiority of SplitMAC over the existing SL frameworks for image classification tasks using the MNIST dataset \cite{MNIST_v2} and the CIFAR-10 dataset \cite{CIFAR10}. Our simulation results indicate that SplitMAC with two-device grouping achieves a significant reduction in the SL latency compared to the existing frameworks. Our results also verify the efficacy of the proposed grouping algorithm in terms of the latency reduction.
        %demonstrate that the use of the two-device grouping strategy provides an additional gain in reducing the latency of SL. 
        %The effectiveness of the proposed grouping algorithm is also verified in terms of the latency reduction.
    \end{itemize}

    %\textcolor{red}{In our preliminary work , we presented only some of analytical results without detailed proofs while developing a device grouping algorithm only for a high-SNR scenario. We extend this preliminary work by offering comprehensive analytical results and developing new device grouping algorithms to overcome the limitation of the algorithm in \cite{SplitMAC_conference}. Specifically, we provide detailed proofs and in-depth discussions to support our analysis on the optimal device grouping. Additionally, we introduce an asymptotically-optimal algorithm and prove its optimality for a low-SNR scenario. We then develop a near-optimal device grouping algorithm that can cover a wide range of SNR  while leveraging the advantages of the other two algorithms. Through this extended research, we not only strengthen the theoretical foundation of the proposed SL framework but also broaden its practical applicability.  }

    The remainder of the paper is organized as follows. 
    In Sec.~II, we first introduce a wireless SL scenario considered in the paper. We then summarize three existing SL approaches and their features. 
    In Sec.~III, we present the proposed SplitMAC framework, which reduces the SL latency by harnessing the nature of wireless multiple access channels. In Sec.~IV, we explore the optimization of the device grouping strategy for the proposed framework. In Sec. V, we extend the proposed framework to make it applicable for practical fading channels and for a general group size.  In Sec. VI, we provide simulation results to verify the superiority of the proposed framework. Finally, in Sec. VII, we present our conclusions and future research directions.

    \section{System Model}\label{Sec:System}
    %\subsection{Split Learning Scenario}
    We consider a wireless SL scenario in which an access point (AP) and multiple wireless devices collaborate to train an AI model. A typical assumption of SL is that training data samples are only possessed by the devices and are not shared with the AP in order to enhance data privacy. Instead, the complete AI model is divided into server-side and device-side models. These models are then trained through collaboration between the AP and the devices during the training process.
    
    %Two models ${\bm w}_{\rm s}$ and ${\bm w}_{\rm d}$ are deployed at the AP and the devices, respectively, and then updated by their collaboration during a training process.
    %To facilitate the training of the AI model without sharing such {\em local} training data samples

    Suppose that the complete AI model is fully represented by its parameters ${\bm w}$.
    Following the aforementioned idea, this model is divided into the server-side and device-side models as follows:
    \begin{align}
        {\bm w} = \{{\bm w}_{\rm d}; {\bm w}_{\rm s}\},
    \end{align}
    where ${\bm w}_{\rm s}$ and ${\bm w}_{\rm d}$ denote the server-side and device-side models, respectively.
    The AP is deployed with the server-side model ${\bm w}_{\rm s}$ and is equipped with a computing server which is capable of training its server-side model. 
    The AP is assumed to have perfect knowledge of the channel conditions of devices. 
    The set of the devices is denoted by $\bar{\mathcal{S}} = \{1, 2, \ldots, N\}$ where $N$ is the total number of the devices. 
    Each device is deployed with the device-side model ${\bm w}_{\rm d}$ and is capable of training this model using its local dataset.
    The local dataset of device $n$ is denoted as $\mathcal{D}_n=\{{\bf z}_{i}, y_i\}_{i=1}^{D_n}$ for all $n\in\bar{\mathcal{S}}$, where ${\bf z}_i$ represents an input data sample, and $y_i$ represents the corresponding label. Then the aggregated dataset over all devices is denoted as $\mathcal{D}=\cup_{n=1}^{N} \mathcal{D}_n$.
    % \begin{itemize}
    %     \item {\bf AP:} The AP is equipped with a computing server that is capable of executing server-side model training. A server-side model, denoted by ${\bm w}_{\rm s}\in \mathbb{R}^{X}$, is deployed at the AP. The AP is assumed to have perfect knowledge of the channel conditions of devices. % device pairing
    %     \item {\bf Device:} We define the set of the devices as $\bar{\mathcal{S}} = \{1, 2, \ldots, N\}$, where $N$ is the total number of the devices. Each device is deployed with a device-side model, denoted by ${\bm w}_{\rm d} \in \mathbb{R}^{X}$. 
    %     Each device is capable of training its device-side model using its local dataset, denoted as $\mathcal{D}_n=\{{\bf z}_{i}, y_i\}_{i=1}^{D_n}, \, \forall n\in\bar{\mathcal{S}}$, where ${\bf z}_i$ represents an input data sample, and $y_i$ is the corresponding label. The aggregated dataset over all devices is denoted as $\mathcal{D}=\cup_{n=1}^{N} \mathcal{D}_n$.
    % \end{itemize}
    Let $\ell ({\bf z}_i, y_i; {\bm w})$ denote the sample-wise loss function, which quantifies the prediction error for a data sample ${\bf z}_i$ in relation to its label $y_i$ given model parameter ${\bm w}$. The average loss function for device $n$ is denoted by $L_n({\bm w})=\frac{1}{|\mathcal{D}_n|}\sum_{\{{\bf z}_i, y_i\}\in \mathcal{D}_n} \ell ({\bf z}_i, y_i; {\bm w})$. The global average loss function is a weighted sum of the losses of the devices where each weight is proportional to the size of each dataset. The global average loss function is expressed as
    \begin{align}
        L({\bm w})=\frac{\sum_{n\in \bar{\mathcal{S}}}|\mathcal{D}_n|L_n({\bm w})}{\sum_{n\in\bar{\mathcal{S}}}|\mathcal{D}_n|}. 
    \end{align}

\begin{figure*}[t]
        \centering
        {\epsfig{file=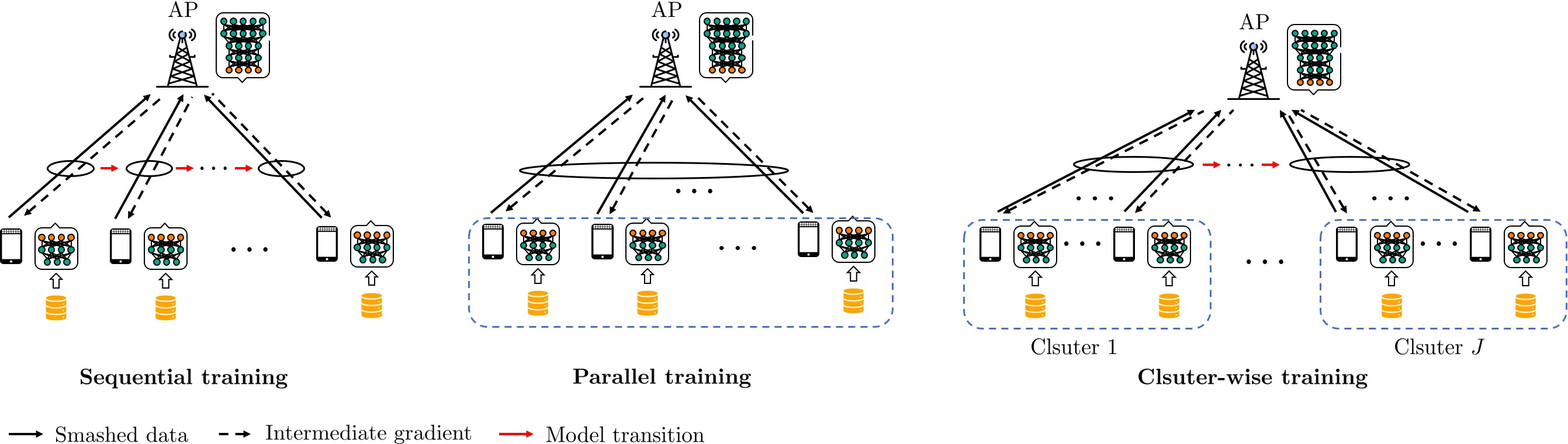,width=17cm}}\vspace{0mm}
        \caption{An illustration of various SL frameworks.}  %\vspace{-3mm}
        \label{fig:various_frameworks}
    \end{figure*}

    The goal of SL is to find the minimizer of $L({\bm w})$. To achieve this goal, three representative training approaches have been developed for SL in the literature. These approaches are summarized below.
    %According to the method to optimize ${\bm w}$, SL frameworks can be divided into the following three types.
    \begin{itemize}
        \item {\bf Sequential training:} In this approach, the model update is executed sequentially over the devices, as illustrated in Fig.~\ref{fig:various_frameworks}. Device $n$ first receives the most recently updated device-side model in learning from the AP and then updates the received model using its local dataset through collaboration with the AP. 
        %by transmitting the smashed data  and intermediate gradient between device $n$ and the AP. 
        % The parameter ${\bm w}$ is updated as the form of
        % \begin{align}\label{eq:GD}
        %     {\bm w} \leftarrow {\bm w} - \eta \frac{|\mathcal{D}_n|}{\sum_{n\in\bar{\mathcal{S}}} |\mathcal{D}_n|} \nabla L_n ({\bm w}),
        % \end{align}
        % by employing the gradient descent algorithm. 
        Then the updated model is conveyed to the next device and the same update process is sequentially executed for all $n\in\Bar{\mathcal{S}}$ in each training round. A major advantage of the sequential training is that it improves the convergence speed of SL by allowing each device to use the most recently updated device-side model.   
        This approach, however, incurs substantial latency for a single training round because device $n$ needs to wait until the end of the training procedures for the $n-1$ previous devices. 
        %for two reasons: (i) sequential device-side model execution and (ii) sequential transmission. 
        %In particular, the uplink communication with the sequential transmission  incurs substantial latency due to the low transmission power of devices.
        
        \item{\bf Parallel training:} 
        %To relieve the latency due to the sequential device-side model execution, the SFL[XX] proposed a parallel SL framework motivated by the FedSGD algorithm. 
        %In this approach, all the devices execute forward propagation in parallel and then transmit the corresponding results, called smashed data, to the AP. Then the AP executes backward propagation and then transmits the corresponding results, called intermediate gradients, to each device. After 
        In this approach, all the devices update their device-side models in parallel, as illustrated in Fig.~\ref{fig:various_frameworks}. Then the AP aggregates these models to construct an updated device-side model. The updated device-side model is distributed to the devices before the beginning of the next training round. The parallel training can reduce the latency of a single training round by allowing parallel processing and update of multiple devices. 
        In this approach, however, the aggregation of the device-side models, which are updated independently, may slow down the convergence speed of SL especially when the dataset is not independent and identically distributed (IID). This phenomenon has been reported in \cite{xiao2020averaging, zhao2018federated} by demonstrating that the heterogeneity of datasets and simple averaging degrade the performance of distributed learning.
        %\textcolor{red}{The heterogeneity of datasets and simple averaging have been shown to degrade the performance of distributed learning, as reported in \cite{xiao2020averaging, zhao2018federated}.} %Also, the relative robustness of sequential SL to this heterogeneity compared to FL was proved through convergence analysis \cite{SL_convergence}
        %While SFL enables parallel device-side training, it still suffers from substantial communication latency. Also, the aggregation of a large number of device-side models may slow down the convergence speed of SL especially when the dataset is non-IID. %[convergence rate of fedavg 랑 용정이형이 알려준 SL learning rate 논문 참조]
        
        \item{\bf Cluster-wise training:} 
        In this approach, devices are divided into multiple clusters, and the model update is executed sequentially across the clusters, as illustrated in Fig.~\ref{fig:various_frameworks}. During the training process for each cluster, devices within the same cluster update their device-side models in parallel. Subsequently, the AP aggregates the models from devices within the same cluster to construct an updated device-side model. Meanwhile, inter-cluster operations follow the sequential training approach, meaning that the updated device-side model in each cluster is passed to the next cluster. 
        The cluster-wise training reduces the latency of SL by utilizing parallel training, while addressing the model aggregation issue by limiting the number of devices in each cluster. 
        Therefore,  the cluster-wise training leverages the advantages of both the sequential and parallel training approaches. 
        % proposed a first-parallel-then-sequential manner framework that reduces the transmission latency while alleviating the model aggregation issue. In CPSL, the AP partition devices into several clusters. Then all devices in the same cluster train their model in parallel using FDMA, and then the global model is updated using the model aggregation over the cluster. The updated model is conveyed to the next cluster and the same process is performed over the cluster. This sequential training over the clusters may relieve the model aggregation issue with the non-IID dataset. First of all, this parallel design reduces both latencies due to the device-side model execution and communication simultaneously.
    \end{itemize}

    % This system model is illustrated in Fig. \ref{fig:system}

    % \begin{figure}[t]
    %     \centering
    %     {\epsfig{file=Figures/system.eps,width=8cm}}\vspace{-3mm}
    %     \caption{The description of the proposed SL scheme}  %\vspace{-3mm}
    %     \label{fig:system}
    % \end{figure}

    \begin{figure*}[t]
        \centering
        {\epsfig{file=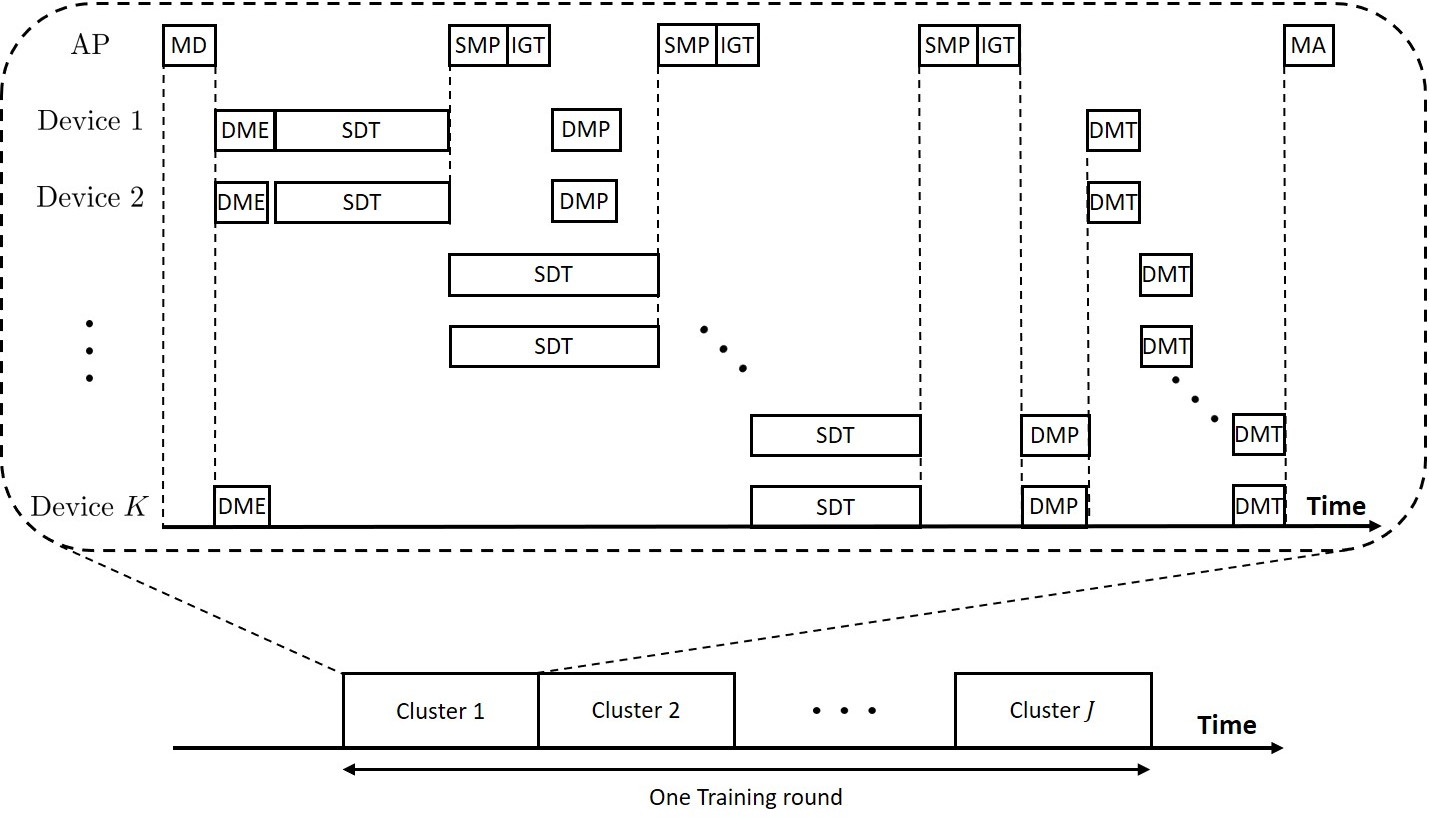,width=17cm}}\vspace{0mm}
        \caption{The procedure of the proposed SplitMAC framework with $L=2$ and $Q=1$.}  %\vspace{-3mm}
        \label{fig:diagram}
    \end{figure*}

    \section{SplitMAC: Split Learning over Multiple Access Channels}
    Inspired by the benefits of cluster-wise training, this section introduces our novel SL framework, referred to as SplitMAC. Our framework builds upon the concept of cluster-wise training but goes a step further in reducing SL latency. It achieves this by leveraging simultaneous transmission over multiple access channels and employing additional latency-reduction strategies.
    In what follows, we outline the key features of the SplitMAC framework and introduce device grouping and clustering considered applied in our framework. We then elaborate on the training procedure for each cluster, followed by the latency analysis of our framework. 

    \subsection{Key Features}
    The key features of the proposed SplitMAC framework are summarized below. 
    \begin{itemize}
        \item {\bf Non-orthogonal multiple success (NOMA):} In SplitMAC, $L$ devices simultaneously access the AP using the uplink NOMA. As a result, these devices transmit their smashed data or device-side model simultaneously over the multiple access channel using the same uplink resources. Then the AP decodes the transmitted data using successive interference cancellation (SIC) which is the capacity-achieving strategy for wireless multiple access channels \cite{takeda2011enhanced}. To this end, the total of $K$ devices is assumed to be divided into the groups of $L$ devices according to a pre-defined device grouping strategy. 
        %\textcolor{red}{ To validate the superiority of NOMA to OMA, we prove that NOMA strictly reduces the latency compared to TDMA using \textbf{Theorem 1}. Additionally, we verify the effectiveness of NOMA compared to FDMA through experiments.}
        
        \item {\bf Local updates of server-side model:} In SplitMAC, the AP updates its model every time it receives smashed data from $QL$ devices (i.e., $Q$ groups of $L$ devices). Consequently, the server-side model can be updated before executing forward/backward propagation for the subsequent devices. This process can be interpreted as a mini-batch update for the server-side model, which can enhance the convergence speed of the server-side model.
        %This allows fast-adaptation of the server-side model because 
        %transmit their smashed data over the multiple access channel, which enables one server-side local update.
        
        \item {\bf Simultaneous uplink-downlink transmission:} 
        In SplitMAC, frequency division duplex (FDD) is adopted to support simultaneous uplink-downlink transmission (i.e., two-way communication). Consequently, the AP can transmit intermediate gradients for the current $L$ devices while simultaneously receiving smashed data from the next $L$ devices. %This implies that the AP does not need to wait for receiving the smashed data from all $K$ devices, in order to start the downlink transmission of the intermediate gradients.
    \end{itemize}    
    %Potential of all these features have never been studied for SL. 
    These features have their own advantages for reducing the total latency of SL, which have never been exploited in the existing FL frameworks.
    Therefore, the proposed SplitMAC framework has a potential to achieve a significant reduction in the SL latency compared to the existing frameworks. 
    
    % \vspace{2mm}
    % \textcolor{red}{{\bf Remark (Comparison to CPSL in \cite{CPSL}):} Even though the CPSL reduces the uplink latency effectively, one inefficiency of the framework stems from the FDMA. If the devices in a cluster transmit their smashed data using FDMA, the AP must wait for all transmissions while staying idle. On the other hand, suppose that the devices communicate with the AP using the TDMA. This approach makes the AP receive the smashed data one by one, which enables the local update on the server. This asynchronous learning with frequent server-side model updates takes advantage of sequential learning. The effectiveness of this core idea will be shown in the following section with the proposed framework.}

    \subsection{Device Grouping and Clustering}
    In the proposed SplitMAC framework, the AP divides a total of $N$ devices into $G$ groups of $L$ devices such that $N=GL$.  
    The devices in the same group are assumed to be \textit{co-scheduled}, implying that these devices utilize the same uplink and downlink resources for communicating with the AP. 
    %Let $G\triangleq K/L$ denote the number of the groups. 
    Let $\mathcal{S}_i \subset \bar{\mathcal{S}}$ represent the index set of $L$ devices in group $i \in \{1,\ldots, G\}$.
    The index sets $\mathcal{S}_1,\ldots,\mathcal{S}_{G}$ are assumed to be mutually exclusive subsets of $\bar{\mathcal{S}}$ such that $\bigcup_{i=1}^G \mathcal{S}_i = \bar{\mathcal{S}}$ and $\mathcal{S}_i \cap \mathcal{S}_j = \emptyset$ for all $i \neq j$. 
    How to determine these subsets will be discussed in Sec.~\ref{Sec:Pairing}.

    After grouping the devices, the AP also constructs clusters, each of which consists of $C$ groups, resulting in each cluster consisting of $K\triangleq CL$ devices. The training procedure of SplitMAC is executed sequentially over the clusters. 
    %Also, the devices in the same cluster download the same device-side model from the AP at the beginning of each training round.  
    Let $J = G/C$ denote the number of the clusters, and let $\mathcal{K}_j$ represent the index set of devices in cluster $j\in\{1,\ldots,J\}$.
    %Then $\mathcal{K}_j$ is constructed as the union of $C$ subsets among $\mathcal{S}_1,\ldots,\mathcal{S}_{G}$. 
    Without loss of generality, we can construct $\mathcal{K}_j$ as $\mathcal{K}_j = \bigcup_{i=(j-1)C+1}^{jC} \mathcal{S}_{i}$ for all $j\in\{1,\ldots,J\}$. 
    It is important to note that the frequency of server-side model updates, $Q$, should not exceed the number of groups in a cluster (i.e., $Q \leq C$). In a system where a cluster consists of $C$ groups, the parameter $Q$ determines the update frequency of the server's model. Specifically, the model on the server is updated each time it receives smashed data from $Q$ groups. It is also noticeable that each device can acquire information on transmission timing and resource allocation by receiving control signals (e.g., downlink control information) from the AP.
    %\textcolor{red}{For each device, the acquisition of resource allocation details and transmission scheduling is enabled by the downlink control information (DCI) located within the physical downlink control channel (PDCCH) as the established protocols in 5G standards.}

    \subsection{In-Cluster Training Procedure}
    %In this stage, model training is executed across clustered devices following a sequence of steps. 
    %For the sake of explanation, we focus on the activities within the $j$-th cluster, denoted by $\mathcal{K}_j$.
    In SplitMAC, device-side and server-side models are updated collaboratively during $T$ training rounds.
    %In each training round, the training procedure is executed sequentially over $J$ clusters. 
    The overall training procedure of SplitMAC with $L=2$ and $Q=1$ in each training round is illustrated in Fig.~\ref{fig:diagram}.
    As can be seen in Fig.~\ref{fig:diagram}, the training procedure is executed sequentially over $J$ clusters, while the training procedure for each cluster involves a total of $8$ steps. 
    %and 
    Details of each step for cluster $j$ in each training round are described below. 
    %Since the same training procedure is executed for the clusters, we will focus only on the procedure for cluster $j$ for the sake of explanation.

    \subsubsection{Step 1 - Device-side model distribution (MD)} 
    In this step, the AP broadcasts an aggregated device-side model, determined by the training procedure of the previous cluster, to all the devices in the current cluster. 
    Then every device in cluster $j$ initializes its device-side model using the aggregated device-side model sent by the AP. 
    The latency of MD is given by 
    \begin{align}
        \tau_{j}^{\rm MD} = \max_{k\in \bar{\mathcal{S}}} \frac{B_{\rm d}}{R_{k}^{\rm DL}},
    \end{align}
    where $B_{\rm d}$ denotes the size (in bits) of the device-side model and $R_k^{\rm DL}$ denotes the downlink transmission rate from the AP to device $k$.
    % Then, each device updates its device-side model using the common device-side model, i.e., 
    % \begin{align}
    %     {\bm w}_{\rm d}^{(j,k)}(t) \leftarrow {\bm w}_{\rm d}^{(j)}(t-1),\, \forall k \in \bar{\mathcal{S}},
    % \end{align}
    % where  ${\bm w}_{\rm d}^{(j-1)}(t-1)$ denotes an aggregated device-side model computed from cluster $j$ and  ${\bm w}_{\rm d}^{(k)}(t)$ denotes the device-side model of device $k$ at training round $t$. 
    The MD step is executed in parallel for the devices within the same cluster at the beginning of the training procedure for each cluster.
    
    \subsubsection{Step 2 - Device-side model execution (DME)} 
    In this step, each device executes forward propagation for its device-side model with respect to every local data sample, generating the corresponding smashed data.
    %up to the cut layer. This propagation is conducted using each device's local data samples. As a result, each device produces its own smashed data. 
    The latency of DME is given by
    \begin{align}
        \tau_{j}^{\rm DME}= \max_{k\in\mathcal{K}_j} \frac{D \gamma_{\rm d}}{f_{k}\kappa_{\rm d}^{\rm F}},
    \end{align}
    where $D$ denotes the batch size, $\gamma_{\rm d}$ denotes the computation workload (in FLOPs) of the forward propagation process of the device-side model for one data sample, $f_{k}$ denotes the central processing unit (CPU) capability of device $k$, and $\kappa_{\rm d}$ denotes the computing intensity (in FLOPs/cycle) of each device. 
    The DME step is executed in parallel for the devices within the same cluster after the reception of the aggregated device-side model sent by the AP.

    \subsubsection{Step 3 - Smashed data transmission (SDT)} In this step, the devices transmit their smashed data, computed in the DME step,  to the AP sequentially over the groups. More specifically, the $L$ devices in group $i$ transmit their smashed data simultaneously over the multiple access channel using the same uplink resources. 
    %which will be decoded by the AP using SIC. 
    Let $B_{\rm sd}$ denote the size (in bits) of the smashed data of device $k$. Then the latency of SDT for group $i$ is given by
    \begin{align}
        \tau_i^{\rm SDT} = \max_{k\in\mathcal{S}_i} \frac{B_{\rm sd}}{R_{k}^{\rm UL}},
    \end{align}
    where $R_k^{\rm UL}$ is the uplink transmission rate from device $k$ to the AP. 
    Upon the reception of the smashed data transmitted from the devices in each group, the AP decodes these smashed data using the capacity-achieving SIC technique \cite{takeda2011enhanced}.
    The SDT step for group $i$ is executed after the completion of the SDT step of group $i-1$, except the special case where group $i$ is the first group of cluster $j$ (i.e., $i=(j-1)C+1$). 
    
    \subsubsection{Step 4 - Server-side model processing (SMP)} In this step, the AP executes the forward/backward propagation, generating the gradient of the server-side model with respect to each smashed data. The AP then updates the server-side model by employing the stochastic gradient descent (SGD) algorithm using the average of the gradients computed for all the smashed data sent by the devices. 
    The AP also generates intermediate gradients that will be transmitted to the devices. Let $f_{\rm s}$ denote the CPU capability of the server. The latency of SMP for each device is given by
    \begin{align}
        \tau^{SMP}=\frac{D(\gamma_{\rm s}^{\rm F}+\gamma_{\rm s}^{\rm B})}{f_{\rm s}\kappa_{\rm s}},
    \end{align}
    where $\gamma_{\rm s}^{\rm F}$ and $\gamma_{\rm s}^{\rm B}$ denote the computation workload (in FLOPs) of the forward and backward propagation processes of the server-side model for one data sample, respectively, and $\kappa_{\rm s}$ denotes the computing intensity (in FLOPs/cycle) of the server.
    The SMP step is executed every time the AP receives the smashed data from $QL$ devices (i.e., $Q$ groups of $L$ devices).

   \subsubsection{Step 5 - Intermediate gradient transmission (IGT)} In this step, the AP transmits the intermediate gradients, computed in the SMP step, to the devices in the same group over the broadcast channel using the same downlink resources. The latency of IGT for group $i$ is given by
    \begin{align}
        \tau_i^{\rm IGT}=\sum_{k \in \mathcal{S}_i} \frac{B_{\rm g}}{R_k^{\rm DL}},
    \end{align}
    where $B_{\rm g}$ denotes the size (in bits) of the intermediate gradients. %Multiple access channel using SIC is not considered here due to the devices' lack of knowledge about the channel conditions. However, 
    By leveraging the two-way communication enabled by FDD, the AP transmits the intermediate gradients for group $i$ via the downlink communication, while simultaneously receiving the smashed data from the devices in group $i+1$ via the uplink communication.  % Nevertheless, the latency caused by IGT is negligible due to the AP's high transmission power.
    The IGT step is executed after the completion of the SMP step. 

    \subsubsection{Step 6 - Device-side model processing (DMP)} In this step, each device in the group executes the backpropagation by using the intermediate gradients sent by the AP and then updates its device-side model. 
    %The update is performed using SGD with its own gradient which is computed from its corresponding intermediate gradient. 
    The latency of DMP for group $i$ is given by 
    \begin{align}
        \tau_i^{\rm DMP}= \max_{k\in\mathcal{S}_i}\frac{D\gamma_{\rm d}^{\rm B}}{f_{k}\kappa_{\rm d}}
    \end{align}
    where $\gamma_{\rm d}^{\rm B}$ denotes the computation workload (in FLOPs) of the back propagation process of the device-side model for one data sample. 
    The DMP step is executed in parallel for devices receiving the intermediate gradients sent by the AP.

    \subsubsection{Step 7 - Device-side model transmission (DMT)} 
    %To enable the AP to perform model aggregation, each updated device-side model must be transmitted to the AP. 
    In this step, each device transmits the weights of the device-side model to the AP in the same way as the SDT step. More precisely, the devices in the same group transmit the weights of their device-side models simultaneously over the uplink multiple access channel. 
    The latency of DMT for group $i$ is given by
    \begin{align}
        \tau_i^{\rm DMT} = \max_{k\in\mathcal{S}_i} \frac{B_{\rm d}}{R_{k}^{\rm UL}}.
    \end{align}
    %The DMP step of the devices in the same group is assumed to be done before the DMT step. 
    This is because both the DMT and SDT steps utilize the same uplink resources for the uplink transmission.
    The DMT step is executed after the completion of all DMP steps in the cluster.
    %st{The DMT step is executed after the completion of the SDT step in each cluster.} \textcolor{red}{}
    
    %It should be noticed that the DMT step for group $1$ cannot start until the end of the SDT step for group $G$.
     %When $K>L$, each device has sufficient spare time for DMP before DMT since the remaining devices in other groups are communicating with the AP for SDT and DMT. Hence, we can neglect the DMP latency for $K>L$.

    \subsubsection{Step 8 - Model aggregation} In this step, the AP collects the device-side models sent from all $K$ devices. The AP then determines the common device-side model, denoted by ${\bm w}_{\rm d}(t)$, as the weighted average of the device-side model, as done in FedSGD \cite{felbab2019optimization}. 
    The corresponding update equation is given by 
        \begin{align}\label{eq:MA}
        {\bm w}_{\rm d}(t) \leftarrow \frac{\sum_{k\in\mathcal{K}_j}|\mathcal{D}_k|{\bm w}_{\rm d}^{(k)}(t)}{\sum_{k\in\mathcal{K}_j}|\mathcal{D}_k|},
    \end{align}
     where ${\bm w}_{\rm d}^{(k)}(t)$ denotes the device-side model of the device $k$. 
    %This aggregation process employs the FedSGD algorithm. 
    %Due to the computational efficiency of FedSGD and the robust CPU capability of the server, the latency incurred during MA is considered negligible.
    %Step~1 to Step 8 will be repeated until both the device-side and server-side models converges. 
    The model aggregation step is executed after the completion of the DMT step in each cluster.

    \begin{table}[t]
    \renewcommand{\arraystretch}{1.2}
    \centering
    \small
    \caption{Definitions of major notations} \label{table:key_params}
    \begin{tabular}{|c | c|}
        \hline
        \bfseries Notation & \bfseries  Definition\\
        \hline \hline
        $L$ & The number of devices in a group\\ \hline
        $C$ & The number of groups in a cluster\\ \hline
        $K$ & The number of devices in a cluster\\ \hline
        $G$ & The number of groups in $\Bar{\mathcal{S}}$ \\ \hline
        $\mathcal{S}_i$  & Set of $L$ devices in group $i$ \\ \hline
        $\mathcal{K}_j$  & Set of $K$ devices in cluster $j$  \\ \hline
        $Q$ & \makecell{Frequency of \\ the server-side model update} \\ \hline
        $D$ & Batch size \\ \hline
        $B_{\rm d}$ & Size of the device-side model \\ \hline
        $B_{\rm sd}$ & Size of the smashed data \\ \hline
        $B_{\rm g}$ & Size of the intermediate gradient \\ \hline
        $B$ & Size of the uplink data \\ \hline
        $R_k^{\rm DL}$ & \makecell{Downlink transmission rate of \\ device $k$} \\ \hline
        $R_k^{\rm UL}$ & \makecell{Uplink transmission rate of \\ device $k$}\\ \hline
        $\gamma_{\rm d}^{\rm F}$ & \makecell{Computation workload of \\ a device for FP} \\ \hline
        $\gamma_{\rm d}^{\rm B}$ &  \makecell{Computation workload of \\ a device for BP}\\ \hline
        $\gamma_{\rm s}^{\rm F}$ & \makecell{Computation workload of \\ the server for FP} \\ \hline
        $\gamma_{\rm s}^{\rm B}$ & \makecell{Computation workload of \\ the server for BP} \\ \hline
        $f_k$ & CPU capability of  device $k$ \\ \hline
        $f_{\rm s}$ & CPU capability of the server \\ \hline
        $\tau(\mathcal{S}_i)$ & Minimum uplink latency for group $\mathcal{S}_i$ \\ \hline
        $f(\{\mathcal{S}_i\})$ & Total uplink latency for grouping $\{\mathcal{S}_i\}$ \\ \hline
    \end{tabular}
\end{table}

    %\subsection{Cluster Transition}
    %After cluster-wise learning is performed on $\mathcal{K}_j$, the learning stage moves to the next pair with the transition of time slot. In this transition, the device-side model is initialized as the recently updated model by the before cluster. Then the cluster-wise learning stage is progressed over the next cluster.
    %The total latency of the proposed SplitMAC framework is illustrated in Fig. \ref{fig:diagram}. 
    \subsection{Latency Analysis}
    We now analyze the total latency of the proposed SplitMAC framework. 
    %\textcolor{red}{The latency for model transition is not handled yet, however, it can be neglected since the information exchanges for model transition are all encapsulated in the In-cluster training process in the form of MD and DMT.} 
    In this analysis, we focus on scenarios\footnote{ These scenarios are not essential for the development or applicability of the proposed framework. In our simulations, we compute the latency even when these scenarios are not valid.} where 
    the sum of latencies of SMP, IGT, and DMP is smaller than the latency of SDT for each group.
    %the sum of the latencies of the SMP and IGT steps is smaller than the latency of the SDT step for each group. 
    This happens when the AP has a sufficiently high computing capability and downlink transmission power.  
    As can be seen in Fig.~\ref{fig:diagram}, the total latency for one training round of SL is given by \eqref{eq:total_latency}. 
    The latency expression in \eqref{eq:total_latency} explicitly shows that the latency is reduced via simultaneous uplink-downlink transmission only when $K>L$.
    It is noteworthy that model transition across different clusters does not require additional latency since the information exchanges for model transition are encapsulated in the form of MD and DMT.
    The definitions of the major notations utilized in this paper are summarized in Table~\ref{table:key_params}.

    % \textcolor{red}{Fig.~\ref{fig:diagram} and the latency expression in \eqref{eq:total_latency} are represented in the scenario where the sum of latencies of SMP, IGT, and DMP is smaller than the latency of SDT for each group. In the simulation, we calculated the latency while reflecting the exception that this assumption is not valid.}
    
    %\textcolor{red}{In fact, there is no guarantee that the sum of DMT latencies exceeds the latency of SMP and IGT for the last group of a cluster. Even though the latency equation \eqref{eq:total_latency} does not reflect this point properly, we handled this exception in simulations.}
    % \begin{figure*}
    % \begin{align}\label{eq:total_latency}
    %     \tau^{\rm tot}=
    %     \begin{cases}
    %         \sum_{j=1}^J \left( \tau_j^{\rm MD} +\tau_j^{\rm DME} +\sum_{i:\mathcal{S}_i\subseteq \mathcal{K}_j} (\tau_i^{\rm SDT}+\tau_i^{\rm DMT}) + L\tau^{SMP} + \tau_{jC}^{\rm IGT} +\tau_{jC}^{\rm DMP} \right), \quad \text{if } K > L, \\
    %         \sum_{j=1}^J \left( \tau_j^{\rm MD} +\tau_j^{\rm DME} +\sum_{i:\mathcal{S}_i\subseteq \mathcal{K}_j} ( \tau_i^{\rm SDT}+ \tau_i^{\rm IGT} + \tau_i^{\rm DMP} + \tau_i^{\rm DMT} ) \right) + N\tau^{\rm SMP}, \quad \text{if } K=L.
    %     \end{cases}
    % \end{align}
    % \hrulefill	
    % \end{figure*}   

    \begin{figure*}
    \begin{align}\label{eq:total_latency}
        \tau^{\rm tot}=
        \begin{cases}
            \sum_{j=1}^J \left( \tau_j^{\rm MD} +\tau_j^{\rm DME} +\sum_{i:\mathcal{S}_i\subseteq \mathcal{K}_j} (\tau_i^{\rm SDT}+\tau_i^{\rm DMT}) + L\tau^{SMP} + \tau_{jC}^{\rm IGT} +\tau_{jC}^{\rm DMP} \right), \quad \text{if } K > L, \\
            \sum_{j=1}^J \left( \tau_j^{\rm MD} +\tau_j^{\rm DME} +\sum_{i:\mathcal{S}_i\subseteq \mathcal{K}_j} ( \tau_i^{\rm SDT}+ \tau_i^{\rm DMT} ) + K\tau^{\rm SMP} + \sum_{i:\mathcal{S}_i\subseteq \mathcal{K}_j} ( \tau_i^{\rm IGT} + \tau_i^{\rm DMP} ) \right) , \quad \text{if } K=L.
        \end{cases}
    \end{align}
    \hrulefill	
    \end{figure*}

    \section{Optimization of Device Grouping for SplitMAC}\label{Sec:Pairing}
    Our analysis in Sec. III-D reveals that the total latency of the proposed SplitMAC framework highly depends on the uplink communication latency required for transmitting the smashed data in the SDT step as well as transmitting the weights of the device-side model in the DMT step. In particular, when the uplink communication bandwidth is limited and the number of participating devices is large, the uplink latency may dominate other types of the latency in SplitMAC.   
    As can be seen in  \cite{noh2021delay} and \cite{sedaghat2018user}, the uplink latency of SplitMAC is closely related to a device grouping strategy which determines how to divide the $N$ devices into the groups of $L$ devices (i.e., determining the sets $\{\mathcal{S}_i\}_{i=1}^G$).
    Motivated by these facts, in this section, we address the problem of optimizing the device grouping of SplitMAC to minimize the uplink communication latency in both the SDT and DMT steps.
    
    %To minimize the overall latency of NOMA-SL presented in Sec. III, it is crucial to minimize the uplink latency of both SDT and DMT procedures, particularly when the uplink communication bandwidth is limited or when the size of the smashed data is large.  

    \subsection{Optimization Problem for Device Grouping}
\begin{figure}[t]
        \centering
        {\epsfig{file=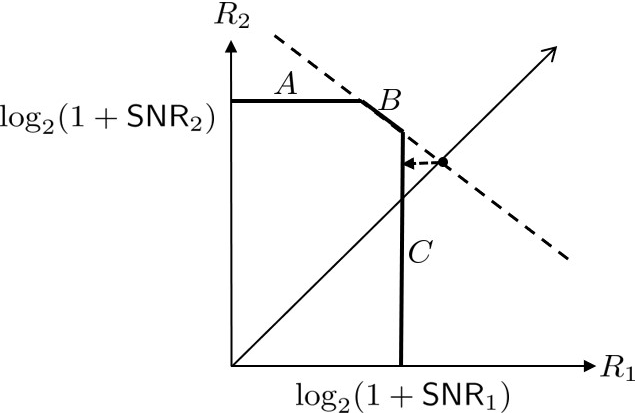,width=8cm}}
        \caption{The derivation of the proposition 3. The solid line is the boundary of $\mathcal{C}(\{1,2\})$, which is composed of three regions $A$, $B$, and $C$. The dashed line is $R_1+R_2=\log_2(1+\textsf{SNR}_1+\textsf{SNR}_2)$ and the outward arrow is $R_1=R_2$. }  %\vspace{-3mm}
        \label{fig:proof}
    \end{figure}

    We start by formulating the problem of finding the optimal device grouping that minimizes the uplink communication latency of SplitMAC.  
    Let $\tau(\mathcal{S}_i)$ denote the minimum latency required for devices in the group $\mathcal{S}_i=\{i_1,i_2,\ldots,i_L\}$ to transmit the smashed data and device-side model. Denote $B$ by the size of the uplink data including smashed data and device-side model (i.e. $B=B_{\rm d}+B_{\rm sd}$). The minimum latency $\tau(\mathcal{S}_i)$ is determined by the straggler which has the longest latency among the devices in $\mathcal{S}_i$. Therefore, the minimum latency $\tau(\mathcal{S}_i)$ for $\mathcal{S}_i$ can be expressed as the solution of a minimax problem as follows:
    \begin{align}\label{eq:P1}
        {\text{(P1)}} \quad \tau(\mathcal{S}_i) = \min_{(R_{i_1},\ldots,R_{i_L}) \in \mathcal{C}(\mathcal{S}_i)}\max_{k\in\mathcal{S}_i} \frac{B}{R_k},
    \end{align}
    where the constraint set $\mathcal{C}(\mathcal{S}_i)$ is the uplink capacity region of $L$ devices in $\mathcal{S}_i$, given by
    \begin{align}\label{eq:capacity_region}
        \mathcal{C}(\mathcal{S}_i)=\bigg\{&(R_{i_1},\ldots, R_{i_L})\bigg| \nonumber\\ &\sum_{k\in \mathcal{S}}R_k \leq \log_2\Big(1+\sum_{k\in\mathcal{S}}\textsf{SNR}_k \Big),\mathcal{S}\subset \mathcal{S}_i  \bigg\}.
    \end{align}
    Here, we denote $R_k$ as $R_k^{\rm UL}$ for the sake of notational simplicity. The capacity region for the case of $L=2$ is depicted as a solid line in Fig. \ref{fig:proof}. The detailed principle of SIC decoding and derivation of the capacity region can be found in \cite{david_tse}.

    Now, define $f(\{\mathcal{S}_i\})$ as the total uplink latency for all device groups given the device grouping in $\{\mathcal{S}_i\}$, given by
    \begin{align}
        f(\{\mathcal{S}_i\})=\sum_{i=1}^G \tau(\mathcal{S}_i).
    \end{align}
    Then the problem of finding the optimal device grouping that minimizes the total uplink latency of SplitMAC can be formulated as follows:
    \begin{align}
        {\text{(P2)}}~~ &\min_{\{\mathcal{S}_i\}}~ f(\{\mathcal{S}_i\}) \nonumber  \\
        &~\text{s.t.}~~ \mathcal{S}_i \cap \mathcal{S}_j = \emptyset,~\cup_{i} \mathcal{S}_i = \bar{\mathcal{S}}.
    \end{align}
    To provide a better understanding for the optimal solution of (P2), we derive the following lemma:
    %To analyze the main problem (P2), we have to write $\tau(\mathcal{S}_i)$ as a closed-form. The following lemmas 1 and 2 enable us to calculate $\tau(\mathcal{S}_i)$.

    \vspace{2mm}
    \begin{lem}\label{lem:1}
        An optimal point $(R_1,R_2,\ldots,R_L)$ which minimizes $\max_k\frac{B}{R_k}$ occurs at the boundary of $\mathcal{C}(\mathcal{S}_i)$.
    \end{lem}

    \begin{IEEEproof}
        Every interior point of $\mathcal{C}(\mathcal{S}_i)$ has a direction which increases every $R_k$ simultaneously. Therefore, the optimal point $(R_1,R_2,\ldots,R_L)$ must occur at the boundary.
    \end{IEEEproof}
    \vspace{2mm}
    
    Although we can limit our analysis to the boundary of $\mathcal{C}(\mathcal{S}_i)$ based on the necessary condition given by {\bf Lemma \ref{lem:1}}, the optimal rate may not always be unique. For example, if an optimal rate exists in either region $A$ or $C$ in Fig. \ref{fig:proof}, such a point would not be considered Pareto optimal because in region $A$, device 1 can increase its transmission rate without adversely affecting the transmission rate of device 2, and a similar conclusion holds for region $C$.
    Therefore, we also derive a sufficient condition for identifying a boundary point as the optimal rate, as given in the following lemma:
    
    \vspace{2mm}
    \begin{lem}\label{lem:2}
        The boundary point ${\bf R}^{\star}=(R_1^{\star},R_2^{\star},\ldots,R_L^{\star})$ such that $R_1=R_2\cdots=R_K$ minimizes \(\max_k\frac{B}{R_k}\).
    \end{lem}

    \begin{IEEEproof}
        If there is a point, ${\bf R}^{\prime}=(R_1^{\prime},R_2^{\prime},\ldots,R_L^{\prime})$, which has smaller \(\max_k \frac{B}{R_k^{\prime}} \) than ${\bf R}^{\star}$, then every $R_k^{\prime}$ must be greater than $R_k^{\star}$. However, there is no such point since ${\bf R}^{\star}$ is a boundary point by {\bf Lemma \ref{lem:1}}.
    \end{IEEEproof}
    \vspace{2mm}
    \textbf{Lemma 1} and \textbf{Lemma 2} can be utilized to derive an optimal rate in the capacity region in \eqref{eq:capacity_region} given group $\mathcal{S}_i$. However, finding the intersection between the line $R_1=\cdots =R_K$ and the boundary of $\mathcal{C}(\mathcal{S}_i)$ is challenging since $\mathcal{C}(\mathcal{S}_i)$ is a polyhedron including $2^L-1$ faces corresponding to subsets of $\mathcal{S}_i$, which requires an exhaustive search.   
    Even if the derivation of the optimal rate is intractable, the results in \textbf{Lemma 1} and \textbf{Lemma 2} suffice to prove the advantage of the device grouping for reducing the uplink latency of SplitMAC, as given in the following theorem:

    \vspace{2mm}
    \begin{thm}\label{thm:TDMA_SIC}
        The proposed SplitMAC framework with device grouping (i.e., $L>1$) always provides a smaller uplink latency compared to the case without device grouping (i.e., $L=1$). In other words, the following inequality holds:
        % In the proposed framework, NOMA always gives a smaller latency than TDMA. In other words, for a given group $\mathcal{S}_i$ with group size $L> 1$, the following inequality holds:
        \begin{align}
            \tau(\mathcal{S}_i)<\sum_{k\in \mathcal{S}_i} \tau(\{k\}).
        \end{align}
    \end{thm}

    \begin{IEEEproof}
        See Appendix~\ref{Apdx:Thm1}.
    \end{IEEEproof}
    \vspace{2mm}

    Our analysis in Theorem~1 indicates that device grouping with $L>1$ is essential for minimizing the uplink latency of SplitMAC, when applying the optimal device grouping determined as the solution of the problem (P2).   
    %the total latency of SplitMAC with device grouping (i.e., $L>1$) is always less than that of the MA-SL without device grouping (i.e., $L=1$).
    Finding the optimal solution of the problem (P2) may involve comparing the latencies of all possible combinations of device grouping for each value of $L$. Such an exhaustive search approach, however, requires tremendous computational complexity and computation time. 
    This implies the need for a fast device grouping algorithm to solve the optimization problem (P2), which is challenging in general due to the complicated entanglement of the transmission rates in the uplink multiple access channel.
    Moreover, when employing the device grouping, the complexity of the SIC-based decoding increases with the number $L$ of devices in each group. This increase in decoding complexity may lead to a significant computational overhead as well as computational time, even for a small number of devices per group.
    Therefore, to circumvent these challenges, in this work, we shall restrict our focus to scenarios where only two devices are grouped together (i.e., $L=2$) and then develop fast device grouping algorithms to determine a sub-optimal but effective solution for the optimal device grouping.
    To emphasize our focus on the case of $L=2$, we will use the term `device pairing' (instead of device grouping) in the rest part of this section.

    \subsection{Proposed Device Pairing Algorithm} 
    For the case of $L=2$, the minimum latency $\tau(\mathcal{S}_i)$ for $\mathcal{S}_i = \{i_1,i_2\}$ can be rewritten as
    %the problem of finding the optimal device pairing that minimizes the uplink latency of SplitMAC can be formulated by simplifying the problem (P2) as follows:  
    \begin{align}\label{eq:P3}
        {\text{(P3)}} \quad \tau(\mathcal{S}_i) = \min_{(R_{i_1},R_{i_2}) \in \mathcal{C}(\mathcal{S}_i)}\max \bigg\{ \frac{B}{R_{i_1}},\frac{B}{R_{i_2}} \bigg\},
    \end{align}
    where the constraint set $\mathcal{C}(\mathcal{S}_i)$ is the uplink capacity region of two devices $i_1$ and $i_2$, given by
    \begin{align}
        \mathcal{C}(\mathcal{S}_i)=\Big\{&(R_{i_1},R_{i_2})\bigg|  R_{i_1}  \leq \log_2\left(1+\textsf{SNR}_{i_1} \right), \nonumber \\ 
        & R_{i_2} \leq \log_2\left(1+\textsf{SNR}_{i_2} \right), \nonumber\\ & R_{i_1} + R_{i_2} \leq \log_2\left(1+\textsf{SNR}_{i_1}+\textsf{SNR}_{i_2} \right)  \Big\}.
    \end{align}
    A minimizer of the problem (P3) is characterized as given in the following theorem:
    %To attain some insights for developing a fast device pairing algorithm, we analyze the characteristics of the optimal device pairing that minimizes the uplink latency of SplitMAC.
    
    \vspace{2mm}
    \begin{prop}\label{prop:closed_form}
        Suppose that $\mathcal{S}_i=\{1,2\}$. A minimizer of the problem (P3) is given by
        \begin{align}\label{eq:optimal_rate}
            R_1&=\min \left\{
            \frac{\log_2(1+\textsf{SNR}_1+\textsf{SNR}_2)}{2},\log_2(1+\textsf{SNR}_1)\right\}, \nonumber\\
            R_2&=\min \left\{
            \frac{\log_2(1+\textsf{SNR}_1+\textsf{SNR}_2)}{2},\log_2(1+\textsf{SNR}_2)\right\}.
        \end{align}
    \end{prop}

    \begin{IEEEproof}
        See Appendix~\ref{Apdx:Prop1}.
    \end{IEEEproof}
    \vspace{2mm}

    Without loss of generality, we set $\textsf{SNR}_{i_1} \leq \textsf{SNR}_{i_2}$. 
    Then, from Theorem~1,  the objective function of the problem (P1) is simplified as 
    \begin{align}
        f(\{\mathcal{S}_i\})=\sum_{i=1}^G\max\bigg\{&\frac{2B}{\log_2(1+{\sf SNR}_{i_1}+{\sf SNR}_{i_2})}, \nonumber \\
        & \frac{B}{\log_2(1+ {\sf SNR}_{i_1})}
        \bigg\}. \label{eq:obj_pair}
    \end{align}

    %To attain some insights for developing a fast device pairing algorithm,
    Now, we further characterize the objective function in \eqref{eq:obj_pair} in two extreme cases: (i) a high SNR scenario (i.e., ${\sf SNR}_k \gg 1$) and (ii) a low SNR scenario (i.e., ${\sf SNR}_k \ll 1$). 
    We then develop two algorithms for determining the optimal device pairing in each scenario. In the remainder of this subsection, we assume that the indices of $N$ devices are sorted in SNR-increasing order, i.e.,
    \begin{align}\label{eq:SNR_sorted}
        \textsf{SNR}_1\leq \textsf{SNR}_2\leq \cdots \leq \textsf{SNR}_N,
    \end{align} 
    for notational simplicity.
    %for two special scenarios where the  \textsf{SNR}s of all the devices are extremely high or extremely low.  
    %we investigate the optimal device pairing for 
    %This particular objective function can be more simplified in scenarios where the overall \textsf{SNR}s of devices become extremely high or extremely low. \textit{Algorithms 1} and \textit{2} are designed to establish optimal pairings in each of these two extreme cases.
    \vspace{2mm}
    \subsubsection{Optimal Device Pairing in High SNR Scenario}
    %Before we propose \textit{algorithm 1}, we define a condition that gives an intuition of \textit{algorithm 1} by lemma 1.
    We first define a condition on the SNRs of the devices, which is useful to characterize the optimal device pairing in the high SNR scenario.
    
    \vspace{1mm}
    \textbf{Definition 1 (Small deviation  condition)}: \textit{The small deviation  condition is said to be satisfied for $\Bar{\mathcal{S}}$ if}
    \begin{align}\label{eq:condition_1}
        \textsf{SNR}_{\max}\leq (1+\textsf{SNR}_{\min})\textsf{SNR}_{\min},
    \end{align}
    \textit{where} $\textsf{SNR}_{\max}$ \textit{and} $\textsf{SNR}_{\min}$ \textit{denote the maximum and minimum SNR among the SNRs of the devices in $\Bar{\mathcal{S}}$, respectively.}
    
    The small deviation condition is likely to hold in high SNR environments because the right-hand term of \eqref{eq:condition_1} includes the square of the SNR. Under this condition, the minimum latency $\tau(\mathcal{S}_i)$ in \eqref{eq:P3} is characterized as a closed-form expression since this condition forces the optimal rates to occur at region $B$ in Fig. \ref{fig:proof}. The result is given in the following lemma:
    %as elaborated in below lemma \ref{lem:lemma_3}.

    \vspace{2mm}
    \begin{lem}\label{lem:lemma_3}
        If the small deviation condition is satisfied for $\Bar{\mathcal{S}}$, the minimum latency $\tau(\mathcal{S}_i)$ in \eqref{eq:P3} is expressed as
        \begin{align}\label{eq:lemma_3}
            \tau(\{i_1,i_2\})=\frac{2B}{\log_2(1+\textsf{SNR}_{i_1}+\textsf{SNR}_{i_2})}, \quad
            \forall i_1\neq i_2.
        \end{align}
    \end{lem}

    \begin{IEEEproof}
     If the small deviation condition is satisfied for $\Bar{\mathcal{S}}$, for any $i_1 \neq i_2$ such that $\textsf{SNR}_{i_1}\leq \textsf{SNR}_{i_2}$,  we have 
    \begin{align}
        &\textsf{SNR}_{i_2}
        \leq (1+\textsf{SNR}_{i_1})\textsf{SNR}_{i_1} \nonumber\\
        &\Leftrightarrow \quad  1+\textsf{SNR}_{i_1}+\textsf{SNR}_{i_2} \leq (1+\textsf{SNR}_{i_1})^2 \nonumber\\
        &\Leftrightarrow \quad {\log_2(1+{\sf SNR}_{i_1})} \geq \frac{\log_2(1+{\sf SNR}_{i_1}+{\sf SNR}_{i_2})}{2}. \nonumber  
    \end{align}
    By Proposition~1 and \eqref{eq:P3}, we obtain the result in \eqref{eq:lemma_3}.
        %See Appendix C.
    \end{IEEEproof}
    \vspace{2mm}
    
    Inspired by the result in {\bf Lemma~\ref{lem:lemma_3}}, we put forth a simple device pairing algorithm to minimize the uplink latency of SplitMAC in the high SNR scenario.
    
    Then, our algorithm pairs the device with the $n$-th largest channel gain with the device with the $n$-th smallest channel gain, i.e., 
    \begin{align}\label{eq:SNR_ordering}
        \mathcal{S}_1=\{1,N\}, \mathcal{S}_2=\{2,N-1\},  \cdots,  \mathcal{S}_{\frac{N}{2}}=\left\{
        \frac{N}{2},\frac{N+1}{2}\right\}.
    \end{align}
    %in a way that a good channel and a bad channel belong to the same pair. 
    We refer to this algorithm as an {\em SNR-balanced pairing} algorithm whose pseudocode is given in {\bf Algorithm~\ref{alg:algorithm_1}}.
    %The design of {\bf Algorithm 1} is guided by the heuristic insight that $f(\{\mathcal{S}_i\})$ is a summation of reciprocals. Nevertheless, this algorithm effectively minimizes $f(\{\mathcal{S}_i\})$ without exhaustive search. Our claim is theoretically substantiated in the following theorem:
    The optimality of {\bf Algorithm 1} is guaranteed in the high SNR scenario (i.e., when the small deviation condition is satisfied). 
    The result is given in the following theorem:

    \vspace{2mm}
    \begin{thm}\label{thm:optimality_1}
    If the small deviation condition is satisfied for $\Bar{\mathcal{S}}$, {\bf Algorithm~1} provides the optimal device pairing, i.e., the solution of the problem (P2) for $L=2$. 
    \end{thm}
    \begin{IEEEproof}	
    	See Appendix~\ref{Apdx:Thm2}.	
    \end{IEEEproof}
    \vspace{2mm}
    
    % can formulate a pairing strategy, $\{\mathcal{S}_i\}$, that aims to minimize the latency function $f(\{\mathcal{S}_i\})$. A specific approach for organizing $\{\mathcal{S}_i\}$ is presented in \textit{Algorithm 1}.
    
    % \textit{Algorithm 1 (SNR-balanced pairing)}: Suppose that $N$ devices are sorted with SNR-increasing order, i.e.
    % \begin{align}
    %     \textsf{SNR}_1\leq \textsf{SNR}_2\leq \cdots \leq \textsf{SNR}_N
    % \end{align}
    
    % Then \textit{algorithm 1} pairs devices into 
    % \begin{align}
    %     \mathcal{S}_1=\{1,N\}, \mathcal{S}_2=\{2,N-1\},  \cdots,  \mathcal{S}_{\frac{N}{2}}=\left\{
    %     \frac{N}{2},\frac{N+1}{2}\right\}
    % \end{align}
    % in a way that a good channel and a bad channel belong to the same pair. A pseudocode of \textit{algorithm 1} is given in {\bf Algorithm~\ref{alg:algorithm_1}}.
    
    \begin{algorithm}[!t]
    	\caption{Pseudocode of the SNR-balanced pairing algorithm}\label{alg:algorithm_1}
    	{\small
    	{\begin{algorithmic}[1]
    		\REQUIRE $\Bar{\mathcal{S}}=\{1,2,\ldots,N\}$
            \ENSURE $\mathcal{S}_1,\mathcal{S}_2,\ldots,\mathcal{S}_{\frac{N}{2}}$
            \FOR {$i=1$ to $\frac{N}{2}$}
			\STATE $\mathcal{S}_{i}=\{i,N-i+1\}$
            \ENDFOR
    	\end{algorithmic}}}
    \end{algorithm}

    \subsubsection{Optimal Device Pairing in Low SNR Scenario}
    Similar to the high SNR scenario, we also define a condition on the SNRs of the devices for deriving the optimal device pairing algorithm in a low SNR scenario. 

    \vspace{1mm}
    \textbf{Definition 2 (Large deviation condition)}: \textit{The large deviation condition is said to be satisfied for $\bar{\mathcal{S}}$ if}
    \begin{align}
        \textsf{SNR}_{i+1}\geq (1+\textsf{SNR}_i) \textsf{SNR}_i,
    \end{align}
    \textit{for all $i \in \{1,2,\ldots,N-1\}$.}
    \vspace{1mm}
    
    The large deviation condition is likely to hold in low SNR environments because $1+\textsf{SNR}_i\approx 1$ for $\textsf{SNR}_i \ll 1$. Under this condition, the minimum latency $\tau(\mathcal{S}_i)$ in \eqref{eq:P3} is characterized as a closed-form expression since this condition forces the optimal rates to occur at region $A$ or $C$ in Fig. \ref{fig:proof}. The result is given in the following lemma:
    %Similar to the small variation condition, this condition also streamlines the expression for latency, $\tau(\mathcal{S}_i)$, as elaborated in below lemma \ref{lem:lemma_4}.

    \vspace{2mm}
    \begin{lem}\label{lem:lemma_4}
        If the large deviation condition is satisfied for $\Bar{\mathcal{S}}$, the minimum latency $\tau(\mathcal{S}_i)$ in \eqref{eq:P3} is expressed as
        \begin{align}\label{eq:lemma_4}
            \tau(\{i_1,i_2\})=\frac{B}{\log_2(1+ \textsf{SNR}_{i_1})}, \quad \forall i_1 < i_2.
        \end{align}
    \end{lem}

    \begin{IEEEproof}
        If the large deviation condition is satisfied for $\Bar{\mathcal{S}}$, we have 
    \begin{align}
        & \textsf{SNR}_{i_2}
        \geq (1+\textsf{SNR}_{i_1})\textsf{SNR}_{i_1}  \nonumber\\
        &\Leftrightarrow \quad  1+\textsf{SNR}_{i_1}+\textsf{SNR}_{i_2} \geq (1+\textsf{SNR}_{i_1})^2 \nonumber\\
        &\Leftrightarrow \quad \frac{\log_2(1+\textsf{SNR}_{i_1}+\textsf{SNR}_{i_2})}{2} \geq \log_2(1+ \textsf{SNR}_{i_1}). \nonumber 
    \end{align}
    By Proposition~1 and \eqref{eq:P3}, we have the result in \eqref{eq:lemma_4}. %This completes the proof since the last inequality is satisfied for all $i_1 <i_2$ by 
        %See Appendix E.
    \end{IEEEproof}
    \vspace{2mm}
    
    Inspired by the result in {\bf Lemma~\ref{lem:lemma_4}}, we introduce a simple device pairing algorithm to minimize the uplink latency of SplitMAC in the low SNR scenario. 
    Under the assumption that the SNRs of the $N$ devices are sorted as in \eqref{eq:SNR_sorted}, our algorithm pairs the device with the $(2n-1)$-th largest channel gain with the $2n$-th largest channel gain for all $n\in\{1,\ldots,N/2\}$, i.e.,
    \begin{align}
        \mathcal{S}_1=\{1,2\}, \mathcal{S}_2=\{3,4\},  \cdots,  \mathcal{S}_{\frac{N}{2}}=\{
        N-1,N\}.
    \end{align}
    We refer to this algorithm as an \textit{SNR-ordered} pairing algorithm whose pseudocode is given in {\bf Algorithm~\ref{alg:algorithm_2}}.
    %The design of \textit{algorithm 2} is guided by the heuristic insight that the latency, $\tau(\{i_1,i_2\})$, given in (\ref{eq:lemma_4}) neglects the effect of $\textsf{SNR}_{i_2}$. \textit{Algorithm 2} also effectively minimizes $f(\{\mathcal{S}_i\})$ without exhaustive search, a claim substantiated by theorem \ref{thm:optimality_2}.
    %The design of {\bf Algorithm~2} is guided by the heuristic insight that $f(\{\mathcal{S}_i\})$ is a summation of reciprocals. Nevertheless, 
    The optimality of {\bf Algorithm~2} is guaranteed in the low SNR scenario (i.e., when the large deviation condition is satisfied). The result is given in the following theorem.
    %s algorithm effectively minimizes $f(\{\mathcal{S}_i\})$ without exhaustive search. Our claim is theoretically proved in the following theorem:
    
    \vspace{2mm}
    \begin{thm}\label{thm:optimality_2}
        If the large deviation condition is satisfied for $\Bar{\mathcal{S}}$, {\bf Algorithm~2} provides the optimal device pairing, i.e., the solution of the problem (P2) for $L=2$. 
    \end{thm}
    \begin{IEEEproof}
        See Appendix~\ref{Apdx:Thm3}.
    \end{IEEEproof}
    \vspace{2mm}
    
    % \textit{Algorithm 2 (SNR-ordered pairing)}: Suppose that $N$ devices are sorted with $\textsf{SNR}$-increasing order.i.e.
    % \begin{align}
    %     \textsf{SNR}_1\leq \textsf{SNR}_2\leq \cdots\leq \textsf{SNR}_N
    % \end{align}
    % Then \textit{algorithm 2} pairs devices into 
    % \begin{align}
    %     \mathcal{S}_1=\{1,2\}, \mathcal{S}_2=\{3,4\},  \cdots,  \mathcal{S}_{\frac{N}{2}}=\{
    %     N-1,N\}
    % \end{align}
    % in a way that similar channels belong to the same pair. A pseudocode of \textit{algorithm 2} is given in \ref{alg:algorithm_2}.

    \begin{algorithm}[!t]
    	\caption{Pseudocode of the SNR-ordered pairing algorithm}\label{alg:algorithm_2}
    	{\small
    	{\begin{algorithmic}[1]
    		\REQUIRE $\Bar{\mathcal{S}}=\{1,2,\ldots,N\}$
            \ENSURE $\mathcal{S}_1,\mathcal{S}_2,\ldots,\mathcal{S}_{\frac{N}{2}}$
            \FOR {$i=1$ to $\frac{N}{2}$}
			\STATE $\mathcal{S}_{i}=\{2i-1,2i\}$
            \ENDFOR
    	\end{algorithmic}}}
    \end{algorithm}

    \subsubsection{Near-Optimal Device Pairing Algorithm} 
    Both the SNR-balanced and SNR-ordered algorithms are proven to be effective for optimizing device pairing under specific channel conditions. 
    However, these algorithms take opposite strategies, and using them in channel environments not suited for each algorithm can lead to performance degradation.
    This fact severely limits the effectiveness of these algorithms under dynamic channel environments with a wide range of SNR. 
    %hinders these algorithms from being applied to a wide range of SNR values. 
    %Proposed \textit{algorithms 1} and \textit{2} achieve the minimum latency under specific \textsf{SNR} conditions. However, the strategies conflict with each other. This conflicting property severely damages the performance of \textit{algorithms 1} and \textit{2} when applied in differing \textsf{SNR} scenarios. This conflicting nature substantially undermines their performance across a wide range of \textsf{SNR} values. 
    To overcome this limitation, we propose a near-optimal algorithm that can cover a wide range of SNR, while leveraging the advantages of both the SNR-balanced and SNR-ordered algorithms. 
    %\textit{algorithm 3} which is robust to the variety of \textsf{SNR}. 
    Recall that the small deviation condition in \eqref{eq:condition_1} is satisfied if and only if
    \begin{align}
        \textsf{SNR}_{\min} \geq \frac{-1+\sqrt{1+4\textsf{SNR}_{\max}}}{2}.
    \end{align}
    Using this property, the proposed algorithm partitions the  set $\bar{\mathcal{S}}$ into the subsets which the small deviation condition is satisfied for. 
    %A precise illustration of the proposed algorithm is described below.
    %\textit{Partitioning step:} 
    Let $\mathcal{G}_k \subset \bar{\mathcal{S}}$ be the $k$-th subset constructed by the proposed algorithm in iteration $k$. 
    Also, let $\Bar{\mathcal{S}}^{(k)}$ be the set of the remaining devices in iteration $k$, where $\Bar{\mathcal{S}}^{(1)}=\Bar{\mathcal{S}}$ and $\Bar{\mathcal{S}}^{(k+1)} = \Bar{\mathcal{S}}^{(k)}\setminus \mathcal{G}_{k}$ for $k\geq 1$. 
    %To this end, the proposed algorithm first initializes $\Bar{\mathcal{S}}^{(1)}=\Bar{\mathcal{S}}$ and then constructs a device subset $\mathcal{G}_k$ from $\Bar{\mathcal{S}}^{(k)}$ in iteration $k$, where 
    % First of all, define $\Bar{\mathcal{S}}^{(1)}=\Bar{\mathcal{S}}$ for initialization. Paritioning step is proceeded by constructing $\mathcal{G}_k$ from $\Bar{\mathcal{S}}^{(k)}$ per each $k$-th iteration.
    %$\Bar{\mathcal{S}}^{(k)} = \Bar{\mathcal{S}}^{(k-1)}\setminus \mathcal{G}_{k-1}$ for $k\geq 2$. 
    %Then the AP constructs a subset $G_k$ by collecting all devices that
    Then the $k$-th subset $\mathcal{G}_k$ is constructed as
    \begin{align}\label{eq:interval}
        \mathcal{G}_k =  \left\{i: \frac{-1+\sqrt{1+4\textsf{SNR}_{\max}^{(k)}}}{2} \leq \textsf{SNR}_i  \leq \textsf{SNR}_{\max}^{(k)} \right\},
    \end{align}
    where $\textsf{SNR}_{\max}^{(k)}$ denotes the maximum value of the SNRs of the devices in $\Bar{\mathcal{S}}^{(k)}$. 
    % \begin{align}\label{eq:interval}
    %     \left\{i: \textsf{SNR}_i \in \left(\frac{-1+\sqrt{1+4\textsf{SNR}_{\max}^{(k)}}}{2}, \textsf{SNR}_{\max}^{(k)} \right)\right\}.
    % \end{align}
    %If the cardinality of $\mathcal{G}_k$ is odd, 
    %subtract the device whose SNR is minimum among $\mathcal{G}_k$ from $\mathcal{G}_k$.  
    If the cardinality of $\mathcal{G}_k$ is odd, the subset $\mathcal{G}_k$ is updated as $\mathcal{G}_k \leftarrow \mathcal{G}_k\setminus \{i_{{\rm min},k}\}$,
    % \begin{align}
    %     \mathcal{G}_k &\leftarrow \mathcal{G}_k\setminus \{i_{{\rm min},k}\}, 
    % \end{align}
    where  $i_{{\rm min},k} = {\argmin}_{i\in \mathcal{G}_k} {\sf SNR}_i$. Note that the length of the interval in \eqref{eq:interval} decreases as the iteration index $k$ increases. 
    %$\textsf{SNR}_{\max}^{(k)}$ decreases with the progress of iteration. 
    The algorithm stops generating the subset $\mathcal{G}_{k}$ at iteration $k$, if there is no device whose SNR lies within the interval \eqref{eq:interval} (i.e., $\mathcal{G}_{k}=\emptyset$), or if the remaining device set is empty  (i.e., $\bar{\mathcal{S}}^{(k)} = \emptyset$).
    % If there is no device whose SNR lies within the interval \eqref{eq:interval} in a certain iteration $k^\prime$, i.e., $\mathcal{G}_{k^\prime}=\emptyset$, the proposed algorithm stops the subset construction.
    % The resulting subsets are given by 
    % %the resulting partition of $\Bar{\mathcal{S}}$ is
    % \begin{align}\label{eq:partition}    \mathcal{G}_1,\mathcal{G}_2,\ldots,\mathcal{G}_{k^\prime -1},
    % \end{align}
    % while the remaining device set is obtained as $\Bar{\mathcal{S}}^{(k^\prime)}$.
    %From the construction, we may assume that the devices in $\Bar{\mathcal{S}}^{(M)}$ have relatively low \textsf{SNR}s. 

    Suppose the proposed near-optimal device pairing algorithm stops generating the subsets at iteration $k^\prime$. Then this algorithm begins pairing the devices in each subset $\mathcal{G}_k$ for $k\in\{1,\ldots,k^\prime-1\}$ independently. Due to the construction, the small deviation condition is satisfied for the devices in every subset. Therefore, the near-optimal algorithm employs the SNR-balanced algorithm's strategy to pair the devices in $\mathcal{G}_k$. If the remaining device set $\bar{\mathcal{S}}^{(k^\prime)}$ is non-empty, the near-optimal  algorithm adopts the SNR-ordered algorithm's strategy to pair the devices in $\bar{\mathcal{S}}^{(k^\prime)}$.
    The rationale behind this choice is that the devices in $\Bar{\mathcal{S}}^{(k^\prime)}$ have relatively low SNR values because the maximum SNR value of the remaining devices in $\Bar{\mathcal{S}}^{(k)}$ decreases with each iteration, i.e., 
    \begin{align}
        \textsf{SNR}_{\max}^{(1)} > \textsf{SNR}_{\max}^{(2)} > \cdots >\textsf{SNR}_{\max}^{(k^\prime)}.
    \end{align}
    The pseudocode of the proposed device pairing algorithm is given in {\bf Algorithm~3}.
    %It completes the procedures of \textit{algorithm 3}. A pseudocode for \textit{algorithm 3} is illustrated in Alg. 3.

    % If $\Bar{\mathcal{S}}$ lies in high \textsf{SNR} regime, then $G_1=\Bar{\mathcal{S}}$ and so \textit{algorithm 3} is same to \textit{algorithm 1}. On the other hand, if $\Bar{\mathcal{S}}$ lies in low \textsf{SNR} regime, then $\bar{\mathcal{S}}^{(M)}=\Bar{\mathcal{S}}$ and so \textit{algorithm 3} is same to \textit{algorithm 2}. This property ensures the robustness of \textit{algorithm 3} to the variety of \textsf{SNR}.

    \begin{algorithm}[!t]
    	\caption{Pseudocode of the proposed near-optimal device pairing algorithm}\label{alg:algorithm_3}
    	{\small
    	{\begin{algorithmic}[1]
    		\REQUIRE $\Bar{\mathcal{S}}=\{1,\ldots,N\}$, $\textsf{SNR}_1,\ldots,\textsf{SNR}_N$ (ascending order)
            \ENSURE $\mathcal{S}_1,\mathcal{S}_2,\ldots,\mathcal{S}_{\frac{N}{2}}$
            \STATE $\Bar{\mathcal{S}}^{(1)}=\Bar{\mathcal{S}}$
            \STATE $k=1$
            \WHILE {${\cal G}_k \neq \emptyset$}
                \STATE Set $\textsf{SNR}_{\max}^{(k)}$ as the maximum \textsf{SNR} among $\Bar{\mathcal{S}}^{(k)}$
                \STATE Set ${\cal G}_{k}$ as (\ref{eq:interval})
                \IF{$|{\cal G}_{k+1}|$ is odd}
                    \STATE ${\cal G}_{k} \leftarrow {\cal G}_{k} \setminus \{\min {\cal G}_{k}\}$
                \ENDIF
                \STATE $\Bar{\mathcal{S}}^{(k+1)}=\Bar{\mathcal{S}}^{(k)}\setminus {\cal G}_{k}$
                \STATE $k\leftarrow k+1$
            \ENDWHILE
            \STATE $M=k-1$
            \FOR {$k=1$ to $k^\prime-1$}
                \STATE Apply {\bf Algorithm 1} by setting $\bar{\mathcal{S}} \leftarrow \mathcal{G}_k$
            \ENDFOR
            \STATE Apply {\bf Algorithm 2} by setting $\bar{\mathcal{S}} \leftarrow \Bar{\mathcal{S}}^{(M)}$
    	\end{algorithmic}}}
    \end{algorithm}

    { \section{Extensions of SplitMAC}
    \subsection{Extension to Practical Fading Channels}
    In Sec. IV, our framework has been predicated on the assumption of static SNR conditions within an SL training procedure. This assumption overlooks the dynamic nature of wireless channels where SNR can vary over time. In this section, we extend our framework to consider practical fading channels where SNR may fluctuate over time.

    To this end, we distinguish the effect of wireless channels into two types: (i) large-scale fading and (ii) small-scale fading. For the large-scale fading, we assume that the large-scale fading coefficients do not vary during an SL training procedure. 
    Under this assumption, we perform our device grouping and clustering methods only once at the beginning of the training procedure based on the large-scale fading coefficients. Subsequently, we utilize the resulting device grouping and clustering throughout the entire training procedure.
    %Consequently, our grouping algorithm now is performed considering only the large-scale fading characteristics, which are assumed to remain static during the training procedure.
    For the small-scale fading, we consider a block-fading channel model in which the wireless channels remain constant within a channel coherence time denoted by $T_{\rm coh}$. Within each coherence time block, we allow the devices within the same group to transmit their data (either smashed data or device-side models) over the MAC at the optimal rate $\mathbf{R}^{\star}$  determined by \textbf{Lemma~2}. This process is repeated until the complete transmission of the data, accommodating the small-scale fading variations during the SDT or DMT steps.

    While the above extension loses the optimality (or near-optimality) of the device grouping under dynamic SNR scenarios, it offers a more realistic and robust framework. 
    %The performance impact of incorporating SNR fluctuations, particularly from small-scale fading, was carefully evaluated. 
    Particularly, there may be a perceived mismatch between grouping based solely on large-scale fading and the actual SNR that also includes small-scale fading. Nevertheless, our numerical analysis indicates that the large-scale fading remains the dominant factor influencing SNR, as will be demonstrated in Sec. VI.  
    %\textcolor{red}{Therefore, while there may be slight adjustments in the grouping algorithm, the overall comparative analysis of different SL frameworks remains valid.}
    
    \subsection{Extension to General Group Size $L$}
    In Sec. IV, we have developed the device pairing algorithms which are applicable only for the special case of $L=2$. To further enhance the  applicability of our algorithm, in this subsection, we develop a general device grouping algorithm applicable for the general case of $L>2$ when $L$ is an even number.

    Suppose that the SNRs of the $N$ devices are sorted in SNR-increasing order without loss of generality. Our general algorithm first divides the whole devices in $\Bar{\mathcal{S}}=\{1,2,\ldots,N\}$ into  $2N/L$ subsets of size $L/2$ as follows:
    \begin{align}\label{eq:algorithm_4_subset}
        \mathcal{H}_1&=\left\{1,2,\ldots,\frac{L}{2}\right\}, \, \mathcal{H}_2=\left\{\frac{L}{2}+1,\ldots,L\right\}, \ldots, \nonumber\\
        \mathcal{H}_{2N/L}&=\left\{N-\frac{L}{2}+1,\ldots,N\right\}.
    \end{align}
    Afterward, define the representative SNRs of each subset as the mean of the SNRs of devices in the subset. Using these representative SNRs, our general algorithm performs  \textbf{Algorithm~3} in the manuscript by treating the $i$-th subset $\mathcal{H}_i$ as device $i$ for $i\in\{1,\ldots,2N/L\}$. Then, apply \textbf{Algorithm~3} as if there are $2N/L$ devices with the SNRs given by $\{\textsf{SNR}^{(1)},\ldots,\textsf{SNR}^{(2N/L)}\}$.
    Let $i_1$ and $i_2$ be the indices of the devices in a {\em virtual} group $i$ determined by  \textbf{Algorithm~3}. Then, an actual group $i$ with size $L$ is determined as the union of $\mathcal{H}_{i_1}$ and $\mathcal{H}_{i_2}$ (i.e., $\mathcal{S}_i = \mathcal{H}_{i_1}\cup \mathcal{H}_{i_2}$).
    The pseudocode of our general algorithm is given in {\bf Algorithm~4}.}
    %Applying \textbf{Algorithm~3} yields $N/L$ unions of two distinct subsets, and these unions correspond to $N/L$ groups of size $L$, which accomplishes the goal of $L$-device grouping. The pseudocode of our new algorithm is given in {\bf Algorithm~4}.}

     \begin{algorithm}[!t]
    	{\caption{Pseudocode of the general device grouping algorithm for even $L$}\label{alg:algorithm_4}
    	{\small
    	{\begin{algorithmic}[1]
    		\REQUIRE $\Bar{\mathcal{S}}=\{1,\ldots,N\}$, $\textsf{SNR}_1,\ldots,\textsf{SNR}_N$ (ascending order)
            \ENSURE $\mathcal{S}_1,\mathcal{S}_2,\ldots,\mathcal{S}_{\frac{N}{L}}$
            \STATE Define subsets $\mathcal{H}_k$ as \eqref{eq:algorithm_4_subset}
            \STATE Define $\textsf{SNR}^{(k)}$ as the mean SNR of $\mathcal{H}_k$
            \STATE Determine $N/L$ groups by applying \textbf{Algorithm 3} with inputs $\Bar{\mathcal{S}}^\prime = \{1,\ldots,2N/L\}$ and $\{\textsf{SNR}^{(1)},\ldots,\textsf{SNR}^{(2N/L)}\}$
            \STATE Set $\mathcal{S}_i$ as the union of two subsets associated with group $i$ for all $i\in\{1,\ldots,N/L\}$ 
    	\end{algorithmic}}}}
    \end{algorithm}
    
    \section{Simulation Results}
    \subsection{Simulation Setup}
    
    In the simulations, we deploy 20 devices uniformly distributed within the cell whose radius is from 10m to 1000m. %\textcolor{red}{The SNR scenarios reflect the fluctuations of small-scale fading as mentioned in Sec. V.} 
    We assume that the uplink and downlink path loss gains of a device whose distance from AP is $d$ (in km) are expressed as follows \cite{channel_model}:
    \begin{align}
            \textrm{PL}^{\rm (dB)} =
            \begin{cases}
                127 + 30 \log_{10}(d)+s, \quad  \textrm{(uplink)},\\
                128.1 + 37.6 \log_{10}(d)+s, \quad \textrm{(downlink)},
            \end{cases}
    \end{align}
    where $s \sim \mathcal{N}(0,\sigma^2)$ encapsulates the shadowing fading, modeled as a Gaussian random variable with zero mean and variance $\sigma^2$. Subsequently, the SNR accounting for only large-scale fading effects is expressed by
    \begin{align}
        \textsf{SNR}^{\rm (large)} = \frac{P_0 \times \textrm{PL}^{\rm (lin)}}{N_0 \times \textrm{BW}},
    \end{align}
    where $P_0$ is the transmission power, $N_0$ is the noise spectral density, and BW is the bandwidth. At the coherence time slot $t$, the instantaneous SNR is 
    \begin{align}
        \textsf{SNR}^{(t)}=\textsf{SNR}^{\rm (large)} \times |h^{(t)}|^2,
    \end{align}
    where $h^{(t)}$ follows a circularly symmetric complex Gaussian distribution $\mathcal{CN}(0,1)$.
    
    While the device grouping is performed using $\textsf{SNR}^{\rm (large)}$ as described in Sec. V-A, the latency is calculated using $\textsf{SNR}^{(t)}$ in our simulations. More specifically, the uplink latency required to transmit data of size $B$ is calculated as follows. First, the size of data that can be transmitted at time slot $t$ is computed as
    \begin{align}
        B^{(t)} = \textrm{BW} \times R^{(t)} \times T_{\rm coh}, 
    \end{align}
    where $R^{(t)}$ is the transmission rate determined by \textbf{Lemma 2}. Then, the number of time slots required to complete the transmission is computed by finding the minimum $T$ such that $\sum_{t=1}^T B^{(t)}\geq B$. Also, the latency for transmitting $B^{(T)}$ is computed as $B^{(T)}/(\textrm{BW}\times R^{(T)})$. As a result, the total uplink time for  transmitting $B$ data is obtained as 
    \begin{align}
        (T-1)\times T_{\rm coh} +\frac{B^{(T)}}{\textrm{BW}\times R^{(T)}}.
    \end{align}

    In the simulations, we set the coherence time to 10 ms. We adopt 4-dB log-normal shadowing variance and Gaussian noise whose power spectral density is given by -174 dBm/Hz \cite{hu2019edge}. The transmit powers of the devices and server are 30 dBm and 42 dBm, respectively, and the bandwidth is 100MHz. The computing capabilities of the devices and the AP are set to $3.4\times10^9$ cycles/s and $100\times 10^9$ cycles/s, respectively. The computing intensities of devices and the server are $\kappa_{\rm d}=4$ (FLOPs/cycle) and $\kappa_{\rm s}=16$ (FLOPs/cycle), respectively. Every parameter is quantized in 32 bits. 

    For the proposed SpitMAC framework, we set $Q=1$ and $L=2$ and adopt the proposed near-optimal algorithm in {\bf Algorithm 3}. As a baseline, we consider three existing distributed learning frameworks: (i) the vanilla SL framework \cite{SL}, referred to as {\em Vanilla SL}, which adopts the sequential-training approach, (ii) a clustering-based SL framework, referred to as {\em Cluster SL}, where the cluster-wise training approach is adopted with FDMA when the subchannel allocation is optimized as proposed in \cite{CPSL}, and (iii) the FL framework \cite{mcmahan17a}, referred to as {\em FedAvg}, where the devices update the entire model by collaborating with the server in an iterative manner. For FedAvg, We assume that the communications between devices and the server are all performed using TDMA.
    We conduct a classification task using the IID and non-IID MNIST datasets. %\textcolor{red}{Instead of fixing the initialization of neural networks, we averaged the latency over random device deployment scenarios.}  
    The batch size is 256, and the learning rate is adjusted so that each scheme shows the fastest convergence. Details of the SL scenarios for each dataset are described below.

    \begin{itemize}
        \item {\bf MNIST:} For the MNIST dataset \cite{MNIST_v2}, we adopt a 12-layer chain topology LeNet \cite{lecun1998gradient} as the global model, which is described in Table \ref{table:1}. The device-side model consists of the first three layers and the server-side model consists of the remaining nine layers. We consider both independent and identically distributed (IID) and non-IID settings. In the non-IID setting, each device possesses training data samples that are exclusively categorized under two types of labels, ensuring that distinct devices contain non-overlapping data samples.
        %For the non-IID setting, each device has training data samples with only two types of labels where distinct devices do not permit overlapping data samples.
        
        \item {\bf CIFAR-10:} For the CIFAR-10 dataset \cite{CIFAR10}, we adopt VGG-16 network pretrained on the ImageNet dataset \cite{VGG} as the global model. The device-side model consists of the first 4 layers and the server-side model consists of the remaining 36 layers. We consider only a non-IID setting which is determined by a Dirichlet distribution with a concentration parameter $\beta=0.3$ \cite{non_IID}.
        
    \end{itemize}    
    
    \begin{table}[t]
    \centering
    \caption{Neural network model considered in our simulations for the MNIST dataset.}
    \begin{tabular}{ cccc } 
    \hline
    \hline
    \textbf{Index} & \textbf{Layer Name} & \textbf{NN Units} & \textbf{Activation}\\
    \hline
    1 & CONV1 & 32, 3 $\times$ 3 & ReLU \\ 
    2 & CONV2 & 32, 3 $\times$ 3 & ReLU \\ 
    3 & POOL1 & 2 $\times$ 2     & None \\
    4 & CONV3 & 64, 3 $\times$ 3 & ReLU \\
    5 & CONV4 & 64, 3 $\times$ 3 & ReLU \\
    6 & POOL2 & 2 $\times$ 2     & None \\
    7 & CONV5 & 128, 3 $\times$ 3 & ReLU \\
    8 & CONV6 & 128, 3 $\times$ 3 & ReLU \\
    9 & POOL3 & 2 $\times$ 2     & None \\
    10 & FC1   & 382              & ReLU \\
    11 & FC2   & 192              & ReLU \\
    12 & FC3   & 10              & Softmax \\
    \hline
    \end{tabular}
    \label{table:1}
    \end{table}

    \subsection{Performance evaluation of SplitMAC}
    \begin{figure}[t]
        \centering
        {\epsfig{file=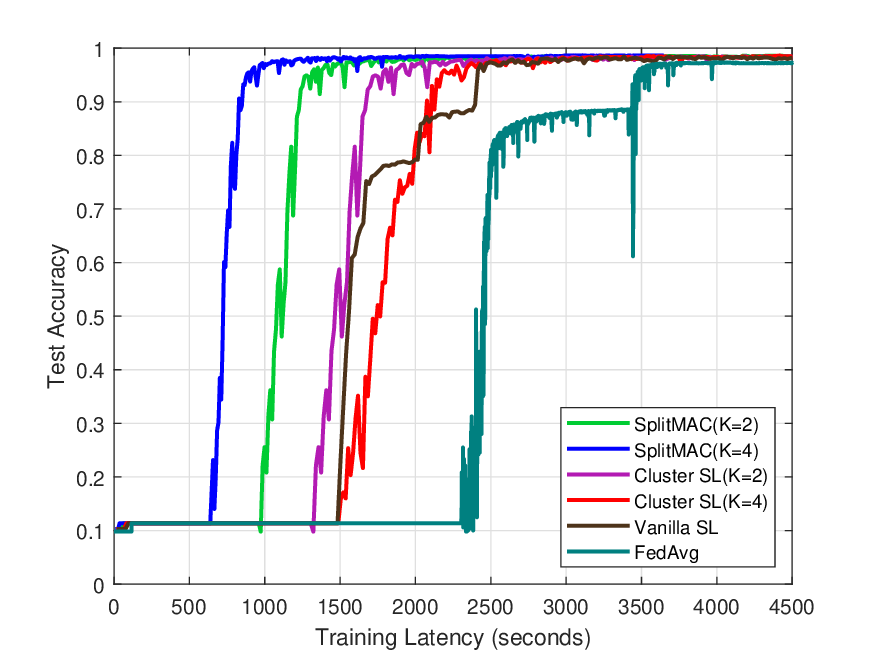,width=8cm}}\vspace{-3mm}
        \caption{Comparison of test accuracy versus training latency for different SL frameworks using the IID MNIST dataset.}  %\vspace{-3mm}
        \label{fig:IID_MNIST}
    \end{figure}

    \begin{figure}[t]
        \centering
        {\epsfig{file=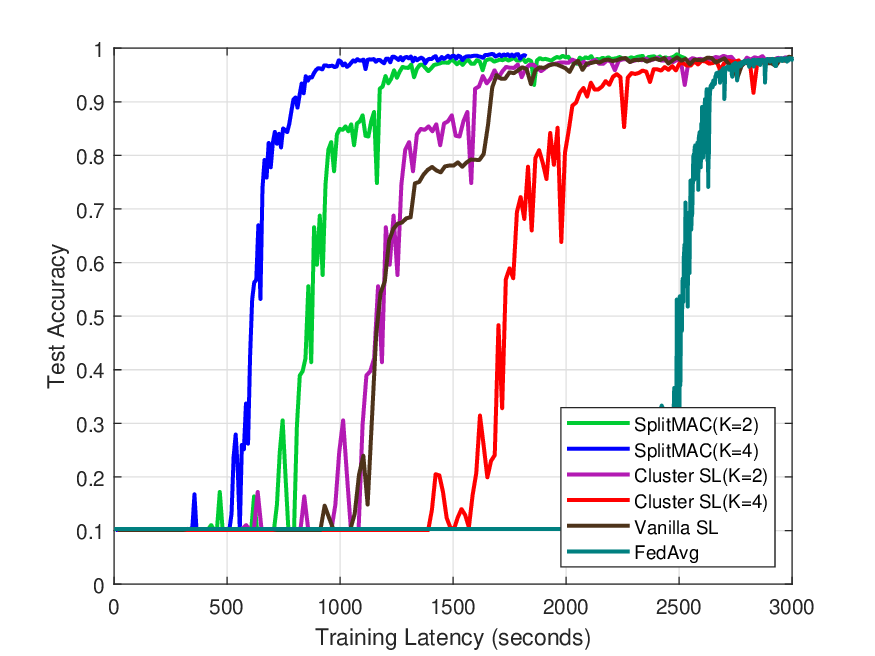,width=8cm}}\vspace{-3mm}
        \caption{Comparison of test accuracy versus training latency for different SL frameworks using the non-IID MNIST dataset.}  %\vspace{-3mm}
        \label{fig:non_IID_MNIST}
    \end{figure}

\begin{figure}[t]
        \centering
        {\epsfig{file=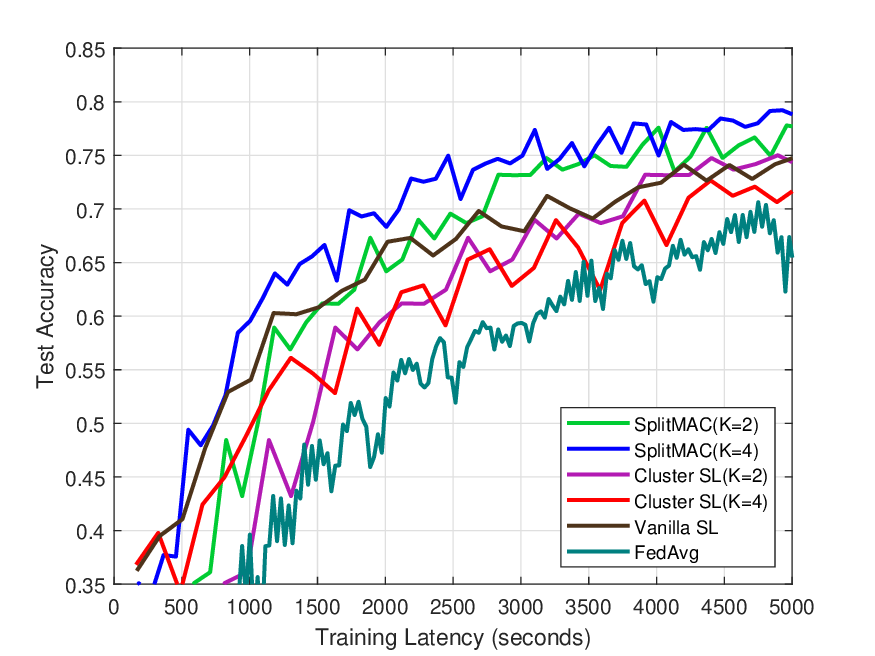,width=8cm}}\vspace{-3mm}
        \caption{Comparison of test accuracy versus training latency for different SL frameworks using the non-IID CIFAR-10 dataset.}  %\vspace{-3mm}
        \label{fig:CIFAR10}
    \end{figure}

    Figs. \ref{fig:IID_MNIST} and \ref{fig:non_IID_MNIST} compare the test accuracy versus training latency for different SL frameworks using the IID and non-IID MNIST datasets, respectively. Similarly, Fig. \ref{fig:CIFAR10} compares the test accuracy versus training latency for different SL frameworks using the non-IID CIFAR-10 dataset. Our simulation results show that SplitMAC with $K=4$ achieves the fastest convergence speed among all the considered SL frameworks, regardless of the datasets and their settings. The convergence speed of FedAvg is inferior to those of the SL frameworks. This result demonstrates the superiority of SL over FL in terms of convergence speed. Contrary to Cluster SL, which experiences a slowdown when $K=4$ compared to $K=2$, SplitMAC sustains its convergence speed through server-side local updates, even demonstrating robustness against the heterogeneity of the MNIST dataset. In addition to this robustness, SplitMAC with $K=4$ gains further latency reduction from the simultaneous uplink-downlink transmission since this feature of SplitMAC can be exploited only for $K>L$ scenario. %\textcolor{red}{This superiority of $K=4$ to $K=2$ stems from the simultaneous uplink-downlink transmission since this property can be exploited only for $K>L$ scenario.} 
    The performance gap between SplitMAC with $K=2$ and Cluster SL with $K=2$ also shows the advantage of utilizing NOMA instead of FDMA, which is one of the key features of SplitMAC, discussed in Sec. III-A.
    
    These results clearly demonstrate the superiority of SplitMAC as a viable solution to reduce the training latency of SL. Our simulation results also show that the clustering-based SL frameworks (i.e., SplitMAC and Cluster SL) provide higher test accuracy than the vanilla SL framework. These results demonstrate that the cluster-wise training approach is more beneficial for reducing the SL latency compared to a sequential training approach. 
    Although both SplitMAC and Cluster SL adopt the cluster-wise training approach, SplitMAC outperforms Cluster SL for the same cluster size $K$.
    This fact implies that the performance gain of SplitMAC is not solely due to the cluster-wise training approach, but also from the  latency-reduction strategies employed in SplitMAC, including simultaneous transmission over multiple access channels.
    It is also notable that the clustering-based SL frameworks converge faster in the IID setting compared to the non-IID setting. This difference arises because aggregating independently-updated device-side models within the same cluster is more effective when the dataset is IID.

\begin{figure}[t]
        \centering
        {\epsfig{file=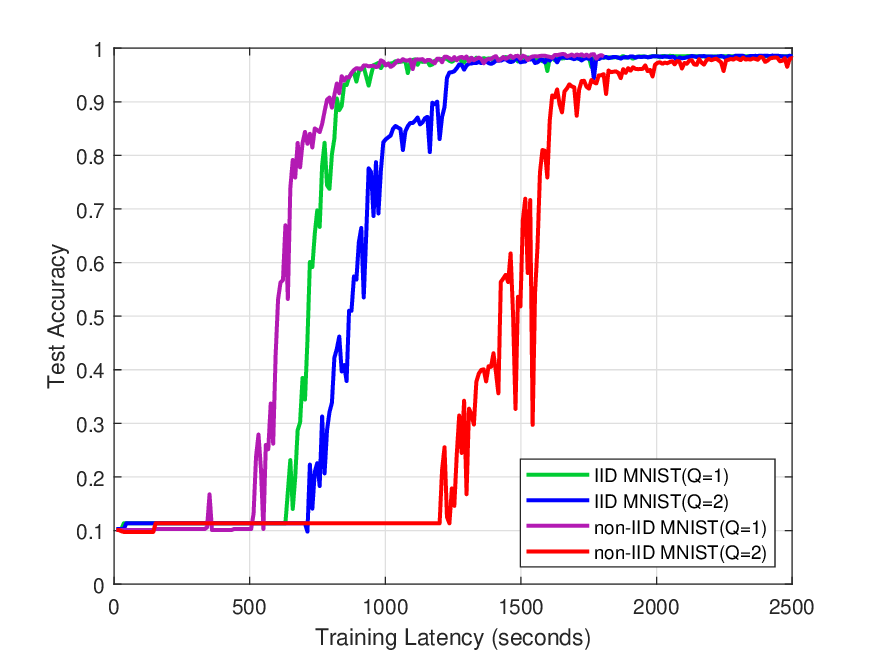,width=8cm}}\vspace{-3mm}
        \caption{Comparison of test accuracy versus training latency for the proposed SplitMAC framework with different values of $Q$ using the IID and non-IID MNIST datasets.}  %\vspace{-3mm}
        \label{fig:impact_of_Q}
    \end{figure}

    Fig. \ref{fig:impact_of_Q} compares the test accuracy versus training latency for the proposed SplitMAC framework with different values of $Q$ using the IID and non-IID MNIST datasets. Fig. \ref{fig:impact_of_Q} shows that SplitMAC with $Q=1$ significantly outperforms SplitMAC with $Q=2$. Recall that the lower the value of $Q$, the more frequently the server-side model is updated. Therefore, our result demonstrates that the frequent local updates of the server-side model effectively improve the convergence speed of SplitMAC, which is one of the key features of our framework discussed in Sec. III-A.
    
    %These results demonstrate that the key features of the proposed framework discussed in Sec. III-A provide significant advantages in reducing the latency of SL. These results also imply that the optimal device grouping algorithm contributes to the reduction in the SL latency. 

\begin{figure}[t]
        \centering
    {\epsfig{file=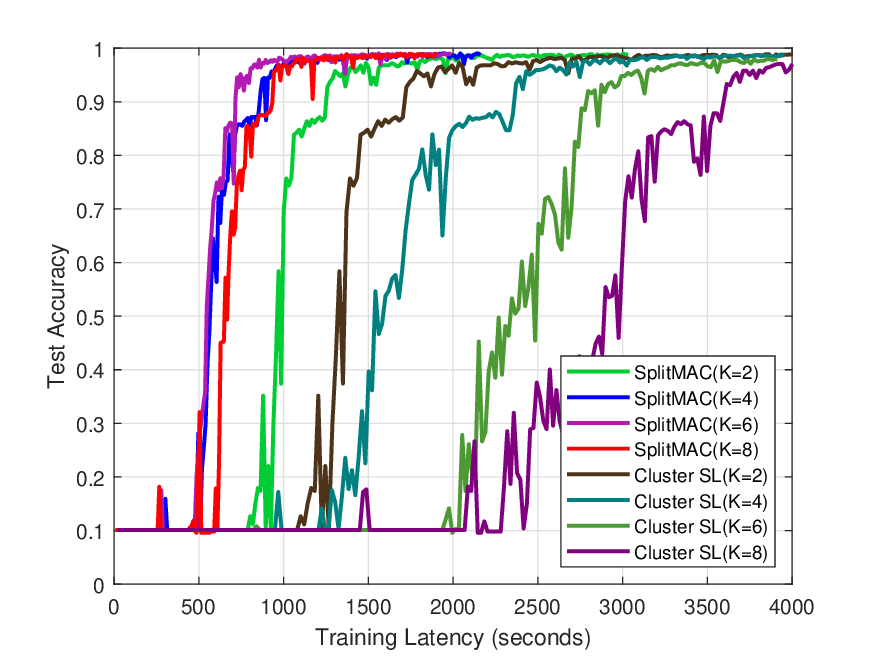,width=8cm}}\vspace{-3mm}
        \caption{Comparison of test accuracy versus training latency for different SL frameworks various cluster sizes using the non-IID dataset.}  %\vspace{-3mm}
        \label{fig:impact_of_K}
    \end{figure}

    Fig. \ref{fig:impact_of_K} compares the test accuracy versus training latency for different SL frameworks with various cluster sizes using the non-IID dataset. To observe the effect of various $K$, we set the number of devices to $N=24$. SplitMAC consistently achieves quicker convergence compared to Cluster SL across various $K$ values. Notably, SplitMAC with $K=6$ exhibits the fastest convergence. Furthermore, while  Cluster SL undergoes the degradation of convergence speed as $K$ increases, the convergence speed of SplitMAC is relatively robust to the change in $K$. This robustness validates that the server-side model update of SplitMAC enhances the model training performance even with large $K$.
    %It is also noteworthy that the differences in convergence speed among various $K$ values become less pronounced in scenarios where $K>L$.

\begin{figure}[t]
        \centering
        {\epsfig{file=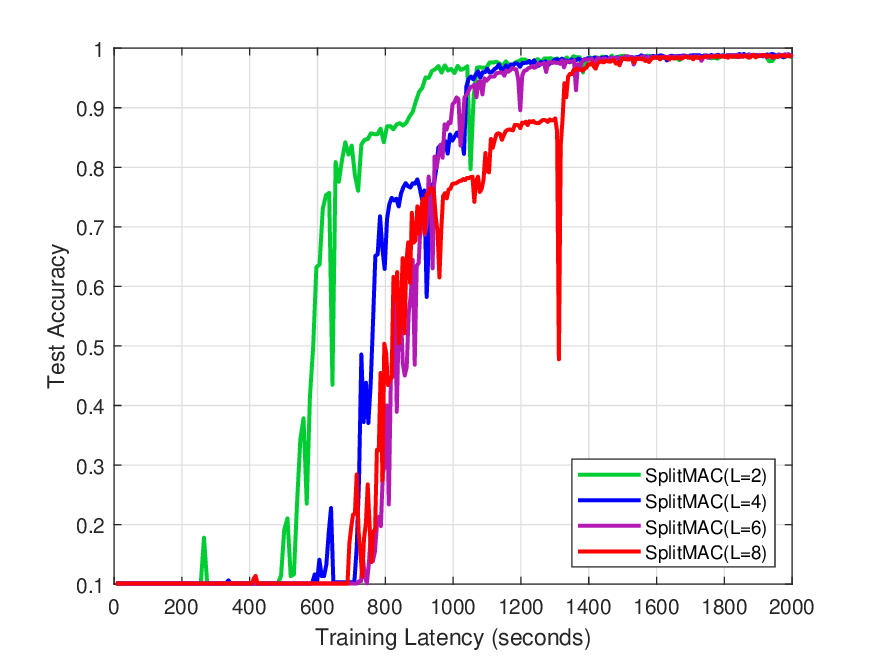,width=8cm}}\vspace{-3mm}
        \caption{Comparison of test accuracy versus training latency  for the proposed SplitMAC framework with different values of $L$ when employing the grouping algorithm in {\bf Algorithm 4}}  %\vspace{-3mm}
        \label{fig:impact_of_L}
    \end{figure}

    Fig. \ref{fig:impact_of_L} compares the test accuracy versus training latency for the proposed SplitMAC framework with different values of $L$ when employing the grouping algorithm in {\bf Algorithm 4}. In this simulation, we set $N=K=24$ and utilize the non-IID MNIST dataset. Fig. \ref{fig:impact_of_L} shows that although there is no dramatic difference in the performances of SplitMAC with various group sizes $L$, the case of $L=2$ provides the best performance among the different values of $L$. It is noteworthy that the complexity of the SIC-based decoding increases with $L$. Therefore, our results demonstrate that the choice of $L=2$ is preferred for SplitMAC in terms of both performance and decoding complexity.
    % shows similar convergence speeds to $L=2$. Nevertheless, from the observation that
    
    \begin{table*}[h]
    \centering
    \caption{Latencies of different training steps in SplitMAC with $K=4$ for one training round.}
    \begin{tabular}{ |c|ccccccc| } 
    \hline
    Dataset& \multicolumn{7}{c|}{MNIST}\\
    \hline
    Step & MD & DME & SDT & SMP & IGT & DMP & DMT\\
    \hline
    Time (s) & 0.18 & 2.81 & {\bf 6.90} & 0.11 & 2.66 & 2.80 & {\bf 0.08}\\ 
    \hline 
    \hline
    Dataset& \multicolumn{7}{c|}{CIFAR-10}\\
    \hline
    Step & MD & DME & SDT & SMP & IGT & DMP & DMT\\
    \hline
    Time (s) & 0.16 & 14.88 & {\bf 72.00} & 3.28 & 27.65 & 14.88 & {\bf 0.18}\\
    \hline
    \end{tabular}
    \label{table:2}
    \end{table*}
    
    % \begin{table}[!ht]
    % \centering
    % \caption{Latencies of different training steps in SplitMAC with $K=4$ for one training round.}
    % \begin{tabular}{ |c|cccc| } 
    % \hline
    % Dataset& \multicolumn{4}{c|}{MNIST}\\
    % \hline
    % Step & SDT & SMP & IGT & DMT\\
    % \hline
    % Time (s)  & {\bf 6.90} & 0.11 & 2.66 & {\bf 0.08}\\ 
    % \hline 
    % \hline
    % Dataset& \multicolumn{4}{c|}{CIFAR-10}\\
    % \hline
    % Step & SDT & SMP & IGT & DMT \\
    % \hline
    % Time (s) & {\bf 72.00} & 3.28 & 27.56 & {\bf 0.18}\\
    % \hline
    % \end{tabular}
    % \label{table:2}
    % \end{table}
    
    We analyze latencies of different training steps in SplitMAC with $K=4$ for one training round using the MNIST and CIFAR-10 datasets. The results are presented in Table~\ref{table:2}. Our analysis reveals that the sum of the latencies for SDT and DMT is the dominant factor in the overall latency. This result validates the significance of reducing the uplink latency in optimizing the overall training latency of SL. Table~\ref{table:2} also shows that the sum of the latencies of the SMP and IGT steps is smaller than the latency of the SDT step. This result confirms our assumption that the AP can transmit the intermediate gradient while simultaneously receiving the next group's smashed data through simultaneous uplink-downlink transmission supported by FDD. In this case, SplitMAC can avoid the latencies of the SMP and IGT steps, which are not negligible, as can be seen in Table~\ref{table:2}. This result demonstrates the effectiveness of the simultaneous uplink-downlink transmission utilized in SplitMAC.
    %\textcolor{blue}{Additionally, our analysis reveals that the sum of the latencies for SDT and DMT is the dominant factor in the overall latency.} This result validates the significance of reducing the uplink latency in optimizing the overall training latency of SL.

    \begin{figure}[t]
        \centering
        {\epsfig{file=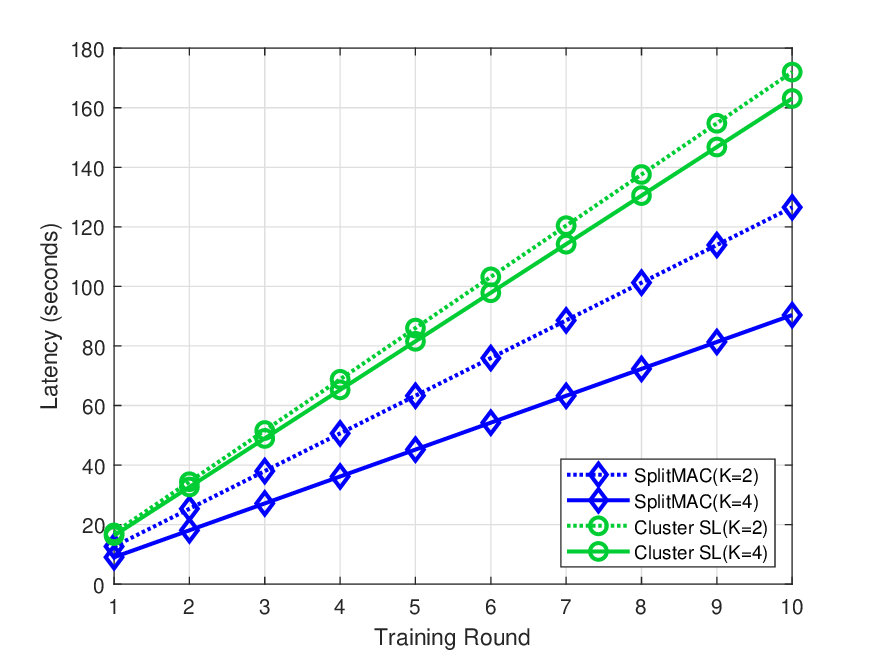,width=8cm}}\vspace{-3mm}
        \caption{Comparison of training latency versus training round for different SL frameworks using the  MNIST dataset.}  %\vspace{-3mm}
        \label{fig:latency_round}
    \end{figure}

    Fig. \ref{fig:latency_round} compares the required latency versus training round for the clustering-based SL frameworks using the MNIST dataset. Fig. \ref{fig:latency_round} shows that SplitMAC explicitly reduces the latency compared to Cluster SL, validating the superiority of simultaneous transmission to Cluster SL in latency reduction. %Even though SplitMAC with $K=2$ has a higher latency than Cluster SL with $K=4$, the test accuracy of the SplitMAC with $K=2$ catches up Cluster SL with $K=4$ as shown in Figs. \ref{fig:IID_MNIST} and \ref{fig:non_IID_MNIST}. These results demonstrate the efficacy of the local updates of the server-side model adopted in SplitMAC.

    \begin{figure}[t]
        \centering
        {\epsfig{file=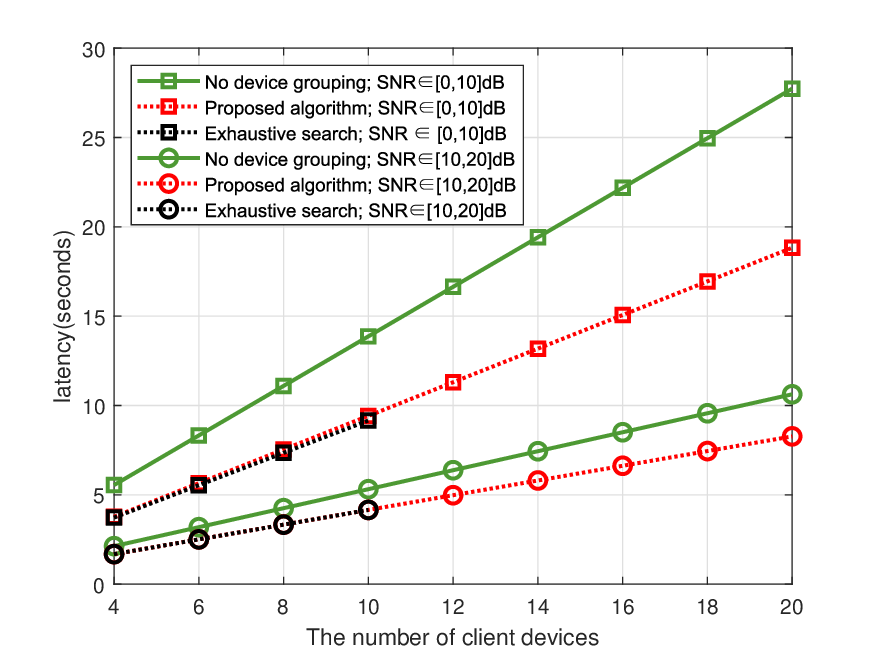,width=8cm}}\vspace{-3mm}
        \caption{Uplink latency versus the number of users for different device grouping strategies using the  MNIST dataset.}  %\vspace{-3mm}
        \label{fig:latency_user}
    \end{figure}
    
    Fig. \ref{fig:latency_user} compares the uplink latency (i.e. the sum of the latencies of the SDT and DMT steps) for one training round versus the total number of devices $N$ for different device grouping strategies. We consider two environments where the SNR is uniformly distributed on $[0,10]$ dB or $[10,20]$ dB. The latency of the exhaustive search approach is only plotted for $N\leq 10$ due to its prohibitive computational complexity when $N>10$. Fig.~\ref{fig:latency_user} shows that the proposed algorithm reduces the uplink latency substantially. Specifically, when $N\leq 10$, the proposed algorithm achieves the optimal latency that can be achieved by the exhaustive search. Furthermore, the proposed algorithm shows remarkable latency reduction for all numbers of devices, especially for $[0,10]$ dB SNR environment. This latency gap is enlarged as the number of users increases.

    \section{Conclusion}
    In this paper, we have proposed a novel SL framework, SplitMAC, designed to reduce the training latency of SL by leveraging simultaneous transmission over multiple access channels. Our SL framework also enhances the test accuracy by enabling the AP to utilize the local update, while reducing the latency via two-way communications supported by FDD. We have analytically demonstrated that the device grouping adopted in the proposed framework contributes to the reduction in the uplink latency of SL. Based on the analysis on transmission rates to minimize the latency, we have devised two asymptotically optimal grouping algorithms and proved their optimality for high-SNR and low-SNR scenarios, respectively. By merging these two algorithms, we have developed the near-optimal grouping algorithm which covers a wide range of SNR. Using simulations, we have demonstrated that SplitMAC achieves superior performance to the existing SL frameworks in the sense of latency reduction. 
    
    % \textcolor{red}{Need to add Future Research Directions... Multiuser extension, Clustering optimization, MIMO}

    %A promising avenue for future research involves developing efficient device grouping algorithms for a general size, where $L>2$. Another promising extension is to leverage multi-antenna techniques for further reducing SL latency. It would also be important to develop an optimal clustering algorithm to minimize the SL latency. 
    
    A promising avenue for future research involves leveraging multi-antenna techniques for further reducing SL latency. Another important direction is to investigate the convergence analysis of the proposed learning method, taking into account the inconsistency between the updates at devices and the server.  Incorporating SL with sensing or task-oriented framework would also be an important direction for future research, as described in \cite{wen2023task}.
    %  Also, the impact of heterogeneous batch sizes among devices on the communication latency and test accuracy needs further investigation. To design grouping further robust to imperfect CSIT, rate-splitting multiple access (RSMA) can be considered an alternative multiple access method.
    \appendices
    \section{Proof of Theorem 1}\label{Apdx:Thm1}

    First, parameterize the optimal $\mathbf{R}^{\star}$ given by {\bf Lemma 2} as $\mathbf{R}^{\star}=(R_1^{\star},R_2^{\star},\ldots,R_L^{\star})=(B\alpha,B\alpha,\ldots,B\alpha)$ in terms of $\alpha$. By {\bf Lemma 1}, $\mathbf{R}^{\star}$ lies on the boundary of $\mathcal{C}(\mathcal{S}_i)$, i.e.,
    \begin{align}
        \sum_{k\in\mathcal{S}^{\prime}}R_k^{\star}=\log_2\left(1+\sum_{k\in\mathcal{S}^{\prime}}\textsf{SNR}_k\right),
    \end{align}
    for some nonempty subset $\mathcal{S}^{\prime}\subseteq \mathcal{S}_i$. Plugging the parametrization into the above condition yields
    \begin{align}
        \alpha = \frac{\log_2\left(1+\sum_{k\in\mathcal{S}^{\prime}}\textsf{SNR}_k\right)}{\sum_{k\in\mathcal{S}^{\prime}}B}.
    \end{align}

    Now, assume that device grouping does not decrease the latency, i.e.,
    \begin{align}\label{eq:alpha}
        \sum_{k\in\mathcal{S}_i}\tau (\{k\}) \leq \tau(\mathcal{S}_i)=\frac{1}{\alpha} = \frac{\sum_{k\in\mathcal{S}^{\prime}}B}{\log_2\left( 1+\sum_{k\in\mathcal{S}^{\prime}}\textsf{SNR}_k\right)}.
    \end{align}
    This assumption gives
    \begin{align}
         \frac{\sum_{k\in\mathcal{S}^{\prime}}B}{\log_2\left( 1+\sum_{k\in\mathcal{S}^{\prime}}\textsf{SNR}_k\right)} 
    & \geq \sum_{k\in\mathcal{S}_i}\tau (\{k\}) \nonumber \\
    & \overset{(a)}{\geq} \sum_{k\in\mathcal{S}^{\prime}}\frac{B}{\log_2(1+\textsf{SNR}_k)} \nonumber \\
    & \geq \frac{\sum_{k\in\mathcal{S}^{\prime}}B}{\log_2\left(1+\max_{k\in\mathcal{S}^{\prime}}\textsf{SNR}_k \right)},
    \end{align}
    where $(a)$ follows from $\mathcal{S}^{\prime}\subseteq \mathcal{S}_i$ and \eqref{eq:P1}.
    % \begin{align}
    %     \frac{\sum_{k\in\mathcal{S}^{\prime}}B}{\log_2\left(1+\max_{k\in\mathcal{S}^{\prime}}\textsf{SNR}_k \right)}
    %     &\leq \sum_{k\in\mathcal{S}^{\prime}}\frac{B}{\log_2(1+\textsf{SNR}_k)}\nonumber\\
    %     &\leq \sum_{k\in\mathcal{S}_i}\tau (\{k\})\nonumber\\
    %     &\leq \tau(\mathcal{S}_i)\nonumber\\
    %     &=\frac{\sum_{k\in\mathcal{S}^{\prime}}B}{\log_2\left( 1+\sum_{k\in\mathcal{S}^{\prime}}\textsf{SNR}_k\right)}.
    % \end{align}
    The above result is a contradiction because 
    \begin{align}
        \max_{k\in\mathcal{S}^{\prime}}\textsf{SNR}_k \geq \sum_{k\in\mathcal{S}^{\prime}}\textsf{SNR}_k,
    \end{align}
    for $|\mathcal{S}^{\prime}|>1$. Additionally, even if $|\mathcal{S}^{\prime}|=1$, the assumption in \eqref{eq:alpha} implies that $\tau(\mathcal{S}_i)=\tau(\mathcal{S}^{\prime})$, which yields
    \begin{align}
        \tau(\mathcal{S}_i)<\tau(\mathcal{S}^{\prime})+\sum_{k\in\mathcal{S}_i\setminus\mathcal{S}^{\prime}}
        \tau(\{k\}) =\sum_{k\in\mathcal{S}_i}\tau(\{k\}).
    \end{align}
    This is also a contradiction to the assumption in \eqref{eq:alpha}.
    Therefore, we come to the conclusion that device grouping always reduces the uplink latency.

	\section{Proof of Proposition 1}\label{Apdx:Prop1}
    We divide the problem (P1) into two cases according to whether the solution of the following equations is feasible or not:
    \begin{align}\label{eq:diag_segment}
            R_1+R_2 = \log_2(1+{\sf SNR}_1+{\sf SNR}_2),~~\text{and}~~
            R_1 = R_2.
        % \begin{cases}
        %     R_1+R_2 &= \log_2(1+{\sf SNR}_1+{\sf SNR}_2),\\
        %     R_1 &= R_2.
        % \end{cases}
    \end{align}
    Let us first consider the case when the solution of \eqref{eq:diag_segment} is feasible, which implies that the point satisfying the condition of {\bf Lemma 2} occurs at the diagonal segment of the boundary of $\mathcal{C}(\mathcal{S}_i)$ (region $B$ in Fig. \ref{fig:proof}). In this case, the optimal rates can be calculated by solving the equations in \eqref{eq:diag_segment}, given by 
    \begin{align}\label{eq:ApdxB:case1}
        R_1&=
        \frac{\log_2(1+\textsf{SNR}_1+\textsf{SNR}_2)}{2}\leq \log_2(1+\textsf{SNR}_1) \nonumber\\
        R_2&=
        \frac{\log_2(1+\textsf{SNR}_1+\textsf{SNR}_2)}{2}\leq \log_2(1+\textsf{SNR}_2).
    \end{align}

    Let us now consider the case  when the solution of  \eqref{eq:diag_segment} is not feasible. Then the optimal rates ${\bf R}^{\star}=(R_1^{\star}, R_2^{\star})$ satisfying $R_1=R_2$ occurs at region $A$ or $C$ in Fig. \ref{fig:proof}. Without loss of generality, suppose that ${\bf R}^{\star}$ occurs at region $C$, i.e., $R_1^{\star}=\log_2(1+\textsf{SNR}_1)$. Since this point is not Pareto-optimal, the rate $R_2$ can be increased without increasing the latency until it meets region $B$. Therefore, one can easily check that
    \begin{align}
        R_2^{\star}\leq \frac{\log_2(1+\textsf{SNR}_1 + \textsf{SNR}_2)}{2} \leq \log_2\left(1+\frac{\textsf{SNR}_2}{1+\textsf{SNR}_1} \right),
    \end{align}
    by the illustration in Fig. \ref{fig:proof}. 
    In this case, projecting the solution of the equations in \eqref{eq:diag_segment} onto region $C$ in Fig. \ref{fig:proof} yields the point $(R_1,R_2)$ expressed as
    \begin{align}\label{eq:ApdxB:case2}
        R_1&=
        \log_2(1+\textsf{SNR}_1) \leq \frac{\log_2(1+\textsf{SNR}_1+\textsf{SNR}_2)}{2}, \nonumber\\
        R_2&=
        \frac{\log_2(1+\textsf{SNR}_1+\textsf{SNR}_2)}{2},
    \end{align}
    implying that the above point gives the same latency with ${\bf R}^{\star}$. An opposite expression is obtained for the case of $R_2^{\star}=\log_2(1+\textsf{SNR}_2)$ due to symmetry. Combining \eqref{eq:ApdxB:case1}, \eqref{eq:ApdxB:case2}, and the opposite expression for \eqref{eq:ApdxB:case2} yields the optimal rate expression in \eqref{eq:optimal_rate}.

    \section{Proof of Theorem 2}\label{Apdx:Thm2}
    By {\bf Lemma 3}, the optimization problem (P2) can be rewritten as follows:
    \begin{align}
        \underset{\{\mathcal{S}_i\}}{\arg\!\min}~& \sum_{i=1}^G \frac{B}{\log_2(1+\textsf{SNR}_{i_1}+\textsf{SNR}_{i_2})} \nonumber \\
        \text{s.t.}~~~ &\mathcal{S}_i \cap \mathcal{S}_j = \emptyset,~\cup_{i} \mathcal{S}_i = \bar{\mathcal{S}} \nonumber\\
        &\mathcal{S}_i=\{i_1,i_2\},~~ i_1<i_2.
    \end{align}
    As the base case, we consider the case of $N=4$. In this case, there are three possible combinations of the device pairing, namely $\{\mathcal{S}_i\}$, $\{\mathcal{T}_i\}$, and $\{\mathcal{U}_i\}$, where $\mathcal{S}_1=\{1,4\}$, $\mathcal{S}_2=\{2,3\}$, $\mathcal{T}_1=\{1,2\}$, $\mathcal{T}_2=\{3,4\}$, $\mathcal{U}_1=\{1,3\}$, and $\mathcal{U}_2=\{2,4\}$.
    % \begin{align}
    %     \mathcal{S}_1&=\{1,4\}, \, \mathcal{S}_2=\{2,3\}, \nonumber \\
    %     \mathcal{T}_1&=\{1,2\}, \, \mathcal{T}_2=\{3,4\}, \nonumber \\
    %     \mathcal{U}_1&=\{1,3\}, \, \mathcal{U}_2=\{2,4\}. \nonumber 
    % \end{align}
    Note that $\{\mathcal{S}_i\}$ is the device pairing obtained from {\bf Algorithm 1}. We first show that $f(\{\mathcal{S}_i\})\leq f(\{\mathcal{T}_i\})$. We start by computing $f(\{\mathcal{T}_i\}) - f(\{\mathcal{S}_i\})$ as follows:
    \begin{align}
        & f(\{\mathcal{T}_i\}) - f(\{\mathcal{S}_i\}) \nonumber \\
        &=\left(\tau(\{1,2\})+\tau(\{3,4\})\right)-\left(\tau(\{1,4\})+\tau(\{2,3\})\right) \nonumber\\
        &=\left(\tau(\{1,2\})-\tau(\{1,4\})\right)-\left(\tau(\{2,3\})-\tau(\{3,4\})\right) \nonumber
        \\
        &=\frac{2B\log_2\left(\frac{1+\textsf{SNR}_1+\textsf{SNR}_4}{1+\textsf{SNR}_1+\textsf{SNR}_2}\right)}{\log_2(1+\textsf{SNR}_1+\textsf{SNR}_2)\log_2(1+\textsf{SNR}_1+\textsf{SNR}_4)} \nonumber\\
        &~~~ -  \frac{2B\log_2\left(\frac{1+\textsf{SNR}_3+\textsf{SNR}_4}{1+\textsf{SNR}_2+\textsf{SNR}_3}\right)}{\log_2(1+\textsf{SNR}_2+\textsf{SNR}_3)\log_2(1+\textsf{SNR}_3+\textsf{SNR}_4)},
    \end{align}
    where the last equality follows from {\bf Lemma 3}. 
    After establishing a common denominator, the resulting numerator is given by  
    %In examining the numerators following the establishment of a common denominator, we find that
    \begin{align}
        &\log_2(1+\textsf{SNR}_2+\textsf{SNR}_3)\log_2(1+\textsf{SNR}_3+\textsf{SNR}_4) \nonumber\\
        &\quad \times \log_2\left(\frac{1+\textsf{SNR}_1+\textsf{SNR}_4}{1+\textsf{SNR}_1+\textsf{SNR}_2}\right) \nonumber\\
        &-\log_2(1+\textsf{SNR}_1+\textsf{SNR}_2)\log_2(1+\textsf{SNR}_1+\textsf{SNR}_4) \nonumber\\
        &\quad \times \log_2\left(\frac{1+\textsf{SNR}_3+\textsf{SNR}_4}{1+\textsf{SNR}_2+\textsf{SNR}_3}\right) \nonumber\\
        &=\log_2(1+\textsf{SNR}_2+\textsf{SNR}_3)\log_2(1+\textsf{SNR}_3+\textsf{SNR}_4) \nonumber\\
        &\quad \times \log_2\left(1+\frac{\textsf{SNR}_4-\textsf{SNR}_2}{1+\textsf{SNR}_1+\textsf{SNR}_2}\right) \nonumber\\
        &-\log_2(1+\textsf{SNR}_1+\textsf{SNR}_2)\log_2(1+\textsf{SNR}_1+\textsf{SNR}_4) \nonumber\\
        &\quad \times \log_2\left(1+\frac{\textsf{SNR}_4-\textsf{SNR}_2}{1+\textsf{SNR}_2+\textsf{SNR}_3}\right) \geq 0
    \end{align}
    where the last inequality follows from $\textsf{SNR}_1\leq \textsf{SNR}_3$. In a similar manner, we can also show that $f(\{\mathcal{U}_i\})-f(\{\mathcal{S}_i\})\geq 0$.
    % \begin{align}
    %     f(\{\mathcal{U}_i\})-f(\{\mathcal{S}_i\})\geq 0.
    % \end{align}
    Therefore, \textbf{Algorithm 1} minimizes the latency for $N=4$.

    Now, suppose that \textbf{Algorithm 1} minimizes the total latency for the case of $N=2M$. To use the induction, let us consider the case of $N=2M+2$. Assume that there is a pairing $\{\mathcal{S}_i^{\prime}\}$ which minimizes the total latency but device 1 is not paired with device $N$, i.e., $\{1,a\},\{b,N\}\in \{\mathcal{S}_i^{\prime}\}$
    % \begin{align}
    %     \{(1,a),(b,N)\}\in\mathcal{S}_i^{\prime}
    % \end{align}
    for some $\textsf{SNR}_a <\textsf{SNR}_N$. Under this assumption, re-pairing them into $\{1,N\}$ and $\{a,b\}$ without changing the rest of the pairs reduces the total latency as shown in the case of  $N=4$. Since it contradicts our original assumption, device 1 must be paired with device $N$. 
    Meanwhile, the remaining devices $2,\ldots,2M-1$ are paired by \textbf{Algorithm 1} as we have already assumed that \textbf{Algorithm 1} minimizes the total latency for the case of $N=2M$.
    Therefore, by induction, \textbf{Algorithm 1} minimizes the total latency under the small variation condition for the general case of $N=2M$ for $M\geq 1$. 

    % \section{Proof of Lemma 4}
    % Note that
    % \begin{align}
    %     &\quad \frac{\log_2(1+\textsf{SNR}_{i_1}+\textsf{SNR}_{i_2})}{2} \geq \log_2(1+\textsf{SNR}_{i_1}) \nonumber\\
    %     &\Leftrightarrow \quad  1+\textsf{SNR}_{i_1}+\textsf{SNR}_{i_2} \geq (1+\textsf{SNR}_{i_1})^2 \nonumber\\
    %     &\Leftrightarrow \quad \textsf{SNR}_{i_2}
    %     \geq (1+\textsf{SNR}_{i_1})\textsf{SNR}_{i_1}.
    % \end{align}
    % This completes the proof since the last inequality is satisfied for all $i_1 <i_2$ by the large difference condition.

    \section{Proof of Theorem 3}\label{Apdx:Thm3}

    By {\bf Lemma 4}, the optimization problem (P2) can be rewritten as follows:
    \begin{align}
        \min_{\{\mathcal{S}_i\}}~ &\sum_{i=1}^G \frac{1}{\log_2(1+\textsf{SNR}_{i_1})} \nonumber\\
        \text{s.t.}~~~ &\mathcal{S}_i \cap \mathcal{S}_j = \emptyset,~\cup_{i} \mathcal{S}_i = \bar{\mathcal{S}} \nonumber\\
        &\mathcal{S}_i=\{i_1,i_2\},~~ i_1<i_2.
    \end{align}
    As the base case, we consider the case of $N=4$. In this case, three possible combinations of the device pairing, namely $\{\mathcal{S}_i\}$, $\{\mathcal{T}_i\}$, and $\{\mathcal{U}_i\}$, where $\mathcal{S}_1=\{1,2\}$, $\mathcal{S}_2=\{3,4\}$, $\mathcal{T}_1=\{1,3\}$, $\mathcal{T}_2=\{2,4\}$, $\mathcal{U}_1=\{1,4\}$, and $\mathcal{U}_2=\{3,4\}$.
    % \begin{align}
    %     \mathcal{S}_1&=\{1,2\}, \, \mathcal{S}_2=\{3,4\}, \nonumber \\
    %     \mathcal{T}_1&=\{1,3\}, \, \mathcal{T}_2=\{2,4\}, \nonumber \\
    %     \mathcal{U}_1&=\{1,4\}, \, \mathcal{U}_2=\{3,4\}.
    % \end{align} 
    Note that $\{\mathcal{S}_i\}$ is the device pairing obtained from {\bf Algorithm 2}. We first show that $f(\{\mathcal{S}_i\})\leq f(\{\mathcal{T}_i\})$ by computing $f(\{\mathcal{T}_i\}) - f(\{\mathcal{S}_i\})$ as follows:
    \begin{align}
        &f(\{\mathcal{T}_i\})-f(\{\mathcal{S}_i\}) \nonumber\\
        &=\left(\tau(\{1,3\})+\tau(\{2,4\})\right)-\left(\tau(\{1,2\})+\tau(\{3,4\})\right) \nonumber\\
        &=\frac{B}{\log_2(1+\textsf{SNR}_2)}-\frac{B}{\log_2(1+\textsf{SNR}_3)}\geq 0,
    \end{align}
    where the inequality follows from  $\textsf{SNR}_2\leq \textsf{SNR}_3$. In a similar manner, we can also show that $f(\{\mathcal{U}_i\})-f(\{\mathcal{S}_i\}) \geq 0$.
    % \begin{align}
    %     f(\{\mathcal{U}_i\})-f(\{\mathcal{S}_i\}) \geq 0.
    % \end{align}
    Therefore, \textbf{Algorithm 2} minimizes the total latency for $N=4$. 
    
    Now, suppose that \textbf{Algorithm 2} minimizes the total latency for the case of $N=2M$. Then we can prove that \textbf{Algorithm 2} is also the optimal device pairing for the case of $N=2M+2$, by using the same re-pairing argument as done in the proof of the Theorem 2. Therefore, \textbf{Algorithm 2} minimizes the total latency under the large variation condition for the general case of $N=2M$ for $M\geq 1$.

    \bibliographystyle{IEEEtran}
    \bibliography{Reference}

\begin{thebibliography}{10}
\providecommand{\url}[1]{#1}
\csname url@samestyle\endcsname
\providecommand{\newblock}{\relax}
\providecommand{\bibinfo}[2]{#2}
\providecommand{\BIBentrySTDinterwordspacing}{\spaceskip=0pt\relax}
\providecommand{\BIBentryALTinterwordstretchfactor}{4}
\providecommand{\BIBentryALTinterwordspacing}{\spaceskip=\fontdimen2\font plus
\BIBentryALTinterwordstretchfactor\fontdimen3\font minus
  \fontdimen4\font\relax}
\providecommand{\BIBforeignlanguage}[2]{{%
\expandafter\ifx\csname l@#1\endcsname\relax
\typeout{** WARNING: IEEEtran.bst: No hyphenation pattern has been}%
\typeout{** loaded for the language `#1'. Using the pattern for}%
\typeout{** the default language instead.}%
\else
\language=\csname l@#1\endcsname
\fi
#2}}
\providecommand{\BIBdecl}{\relax}
\BIBdecl

\bibitem{Konecny:15}
J.~Kone{\v{c}}n{\`y}, B.~McMahan, and D.~Ramage, ``Federated optimization:
  Distributed optimization beyond the datacente,'' 2015,
  \textit{arXiv:1511.03575}.

\bibitem{mcmahan17a}
B.~McMahan, E.~Moore, D.~Ramage, S.~Hampson, and B.~A.~y. Arcas,
  ``Communication-efficient learning of deep networks from decentralized
  data,'' in \emph{Proc. Int. Conf. Artificial Intell. Statist. (AISTATS)},
  Fort Lauderdale, FL, USA, Apr. 2017, pp. 1273--1282.

\bibitem{niknam2020federated}
S.~Niknam, H.~S. Dhillon, and J.~H. Reed, ``Federated learning for wireless
  communications: Motivation, opportunities, and challenges,'' \emph{IEEE
  Commun. Mag.}, vol.~58, no.~6, pp. 46--51, Jun. 2020.

\bibitem{jeon2020compressive}
Y.-S. Jeon, M.~M. Amiri, J.~Li, and H.~V. Poor, ``A compressive sensing
  approach for federated learning over massive {MIMO} communication systems,''
  \emph{IEEE Trans. Wireless Commun.}, vol.~20, no.~3, pp. 1990--2004, Mar.
  2021.

\bibitem{FedSQCS}
Y.~Oh, N.~Lee, Y.-S. Jeon, and H.~V. Poor, ``Communication-efficient federated
  learning via quantized compressed sensing,'' \emph{IEEE Trans. Wireless
  Commun.}, vol.~22, no.~2, pp. 1087--1100, Feb. 2023.

\bibitem{SL}
O.~Gupta and R.~Raskar, ``Distributed learning of deep neural network over
  multiple agents,'' \emph{J. Netw. Comput. Appl.}, vol. 116, pp. 1--8, Aug.
  2018.

\bibitem{vepakomma2018split}
P.~Vepakomma, O.~Gupta, T.~Swedish, and R.~Raskar, ``Split learning for health:
  Distributed deep learning without sharing raw patient data,'' in \emph{Proc.
  Int. Conf. Learn. Represent. (ICLR) Workshop AI Social Good}, New Orleans,
  LA, USA, May 2019, pp. 1--7.

\bibitem{VFL}
K.~Wei, J.~Li, C.~Ma, M.~Ding, S.~Wei, F.~Wu, G.~Chen, and T.~Ranbaduge,
  ``Vertical federated learning: Challenges, methodologies and experiments,''
  2022, \textit{arXiv:2202.04309}.

\bibitem{letaief2021edge}
K.~B. Letaief, Y.~Shi, J.~Lu, and J.~Lu, ``Edge artificial intelligence for 6g:
  Vision, enabling technologies, and applications,'' \emph{IEEE J. Sel. Areas
  Commun.}, vol.~40, no.~1, pp. 5--36, Jan. 2022.

\bibitem{tran2022privacy}
N.-P. Tran, N.-N. Dao, T.-V. Nguyen, and S.~Cho, ``Privacy-preserving learning
  models for communication: A tutorial on advanced split learning,'' in
  \emph{Proc. Int. Conf. Inf. Commun. Technol. Convergence (ICTC)}, Jeju
  Island, Republic of Korea, Oct. 2022, pp. 1059--1064.

\bibitem{SL_FL_comparison}
A.~Singh, P.~Vepakomma, O.~Gupta, and R.~Raskar, ``Detailed comparison of
  communication efficiency of split learning and federated learning,'' 2019,
  \textit{arXiv:1909.09145}.

\bibitem{SL_convergence}
Y.~Li and X.~Lyu, ``Convergence analysis of sequential federated learning on
  heterogeneous data,'' in \emph{Adv. Neural Inf. Process. Syst. (NeurIPS)},
  New Orleans, LA, USA, Dec. 2023, pp. 56\,700--56\,755.

\bibitem{shiranthika2023decentralized}
C.~Shiranthika, P.~Saeedi, and I.~V. Baji{\'c}, ``Decentralized learning in
  healthcare: A review of emerging techniques,'' \emph{IEEE Access}, vol.~11,
  pp. 54,188--54,209, Jun. 2023.

\bibitem{SL_SQ_1}
B.~Yuan, S.~Ge, and W.~Xing, ``A federated learning framework for healthcare
  {IoT} devices,'' 2020, \textit{arXiv:2005.05083}.

\bibitem{SL_SQ_2}
J.~Wang, H.~Qi, A.~S. Rawat, S.~Reddi, S.~Waghmare, F.~X. Yu, and G.~Joshi,
  ``Fed{L}ite: A scalable approach for federated learning on
  resource-constrained clients,'' 2022, \textit{arXiv:2201.11865}.

\bibitem{oh2023communication}
Y.~Oh, J.~Lee, C.~G. Brinton, and Y.-S. Jeon, ``Communication-efficient split
  learning via adaptive feature-wise compression,'' 2023,
  \textit{arXiv:2307.10805}.

\bibitem{thapa2022splitfed}
C.~Thapa, P.~C.~M. Arachchige, S.~Camtepe, and L.~Sun, ``Split{F}ed: When
  federated learning meets split learning,'' in \emph{Proc. AAAI Conf. Artif.
  Intell.}, Jun. 2022, pp. 8485--8493.

\bibitem{CPSL}
W.~Wu, M.~Li, K.~Qu, C.~Zhou, X.~Shen, W.~Zhuang, X.~Li, and W.~Shi, ``Split
  learning over wireless networks: Parallel design and resource management,''
  \emph{IEEE J. Sel. Areas Commun.}, vol.~41, no.~4, pp. 1051--1066, Apr. 2023.

\bibitem{EPSL}
Z.~Lin, G.~Zhu, Y.~Deng, X.~Chen, Y.~Gao, K.~Huang, and Y.~Fang, ``Efficient
  parallel split learning over resource-constrained wireless edge networks,''
  2023, \textit{arXiv:2303.15991}.

\bibitem{chen2017exploiting}
X.~Chen, Z.~Zhang, C.~Zhong, and D.~W.~K. Ng, ``Exploiting multiple-antenna
  techniques for non-orthogonal multiple access,'' \emph{IEEE J. Sel. Areas
  Commun.}, vol.~35, no.~10, pp. 2207--2220, Jul. 2017.

\bibitem{vaezi2019interplay}
M.~Vaezi, G.~A.~A. Baduge, Y.~Liu, A.~Arafa, F.~Fang, and Z.~Ding, ``Interplay
  between {NOMA} and other emerging technologies: A survey,'' \emph{IEEE Trans.
  Cogn. Commun. Netw.}, vol.~5, no.~4, pp. 900--919, Aug. 2019.

\bibitem{wei2019performance}
Z.~Wei, L.~Yang, D.~W.~K. Ng, J.~Yuan, and L.~Hanzo, ``On the performance gain
  of {NOMA} over {OMA} in uplink communication systems,'' \emph{IEEE Trans.
  Wireless Commun.}, vol.~68, no.~1, pp. 536--568, Oct. 2019.

\bibitem{chen2017optimization}
Z.~Chen, Z.~Ding, X.~Dai, and R.~Zhang, ``An optimization perspective of the
  superiority of {NOMA} compared to conventional {OMA},'' \emph{IEEE Trans.
  Signal Process.}, vol.~65, no.~17, pp. 5191--5202, 2017.

\bibitem{task_offloading}
Z.~Song, Y.~Liu, and X.~Sun, ``Joint task offloading and resource allocation
  for {NOMA}-enabled multi-access mobile edge computing,'' \emph{IEEE Trans.
  Commun.}, vol.~69, no.~3, pp. 1548--1564, 2021.

\bibitem{noh2021delay}
V.~D. Tuong, W.~Noh, and S.~Cho, ``Delay minimization for {NOMA}-enabled mobile
  edge computing in industrial internet of things,'' \emph{IEEE Trans. Ind.
  Informat.}, vol.~18, no.~10, pp. 7321--7331, Oct. 2021.

\bibitem{MNIST_v2}
Y.~Lecun, L.~Bottou, Y.~Bengio, and P.~Haffner, ``Gradient-based learning
  applied to document recognition,'' \emph{Proc. IEEE}, vol.~86, no.~11, pp.
  2278--2324, Nov. 1998.

\bibitem{CIFAR10}
A.~Krizhevsky and G.~Hinton, ``Learning multiple layers of features from tiny
  images,'' M.S. thesis, Univ. Toronto, Toronto, ON, Canada, 2009.

\bibitem{xiao2020averaging}
P.~Xiao, S.~Cheng, V.~Stankovic, and D.~Vukobratovic, ``Averaging is probably
  not the optimum way of aggregating parameters in federated learning,''
  \emph{Entropy}, vol.~22, no.~3, p. 314, 2020.

\bibitem{zhao2018federated}
Y.~Zhao, M.~Li, L.~Lai, N.~Suda, D.~Civin, and V.~Chandra, ``Federated learning
  with non-iid data,'' 2018, \textit{arXiv:1806.00582}.

\bibitem{takeda2011enhanced}
T.~Takeda and K.~Higuchi, ``Enhanced user fairness using non-orthogonal access
  with sic in cellular uplink,'' in \emph{Proc. IEEE Veh. Technol. Conf.
  (VTC)}, San Francisco, CA, USA, Sep. 2011, pp. 1--5.

\bibitem{felbab2019optimization}
V.~Felbab, P.~Kiss, and T.~Horv{\'a}th, ``Optimization in federated learning,''
  in \emph{Proc. 19th Conf. Inf. Technol.–Appl. Theory (ITAT)}, Donovaly,
  Slovakia, Sep 2019, pp. 58--65.

\bibitem{sedaghat2018user}
M.~A. Sedaghat and R.~R. M{\"u}ller, ``On user pairing in uplink {NOMA},''
  \emph{IEEE Trans. Wireless Commun.}, vol.~17, no.~5, pp. 3474--3486, Mar.
  2018.

\bibitem{david_tse}
T.~David and P.~Viswanath, ``{Fundamentals of Wireless Communication},'' Univ.
  Cambridge, Cambridge, Press, U.K., 2005.

\bibitem{channel_model}
J.~Zhang, M.~You, G.~Zheng, I.~Krikidis, and L.~Zhao, ``Model-driven learning
  for generic {MIMO} downlink beamforming with uplink channel information,''
  \emph{IEEE Trans. Wireless Commun.}, vol.~21, no.~4, pp. 2368--2382, Apr.
  2022.

\bibitem{hu2019edge}
X.~Hu, L.~Wang, K.-K. Wong, M.~Tao, Y.~Zhang, and Z.~Zheng, ``Edge and central
  cloud computing: A perfect pairing for high energy efficiency and
  low-latency,'' \emph{IEEE Trans. Wireless Commun.}, vol.~19, no.~2, pp.
  1070--1083, Feb. 2019.

\bibitem{lecun1998gradient}
Y.~LeCun, L.~Bottou, Y.~Bengio, and P.~Haffner, ``Gradient-based learning
  applied to document recognition,'' \emph{Proc. IEEE}, vol.~86, no.~11, pp.
  2278--2324, Nov 1998.

\bibitem{VGG}
K.~Simonyan and A.~Zisserman, ``Very deep convolutional networks for
  large-scale image recognition,'' in \emph{Proc. Int. Conf. Learn. Represent.
  (ICLR)}, San Diego, CA, USA, May 2015, pp. 1--14.

\bibitem{non_IID}
Q.~Li, Y.~Diao, Q.~Chen, and B.~He, ``Federated learning on non-iid data silos:
  An experimental study,'' in \emph{Proc. IEEE Int. Conf. Data Eng. (ICDE)},
  Kuala Lumpur, Malaysia, May 2022, pp. 965--978.

\bibitem{wen2023task}
D.~Wen, P.~Liu, G.~Zhu, Y.~Shi, J.~Xu, Y.~C. Eldar, and S.~Cui, ``Task-oriented
  sensing, computation, and communication integration for multi-device edge
  {AI},'' \emph{IEEE Trans. Wireless Commun.}, early access, Aug. 14, 2023,
  doi: 10.1109/TWC.2023.3303232.

\end{thebibliography}
    
\end{document}